\documentclass[11pt,a4paper]{article}

\usepackage{jheppub}
\usepackage{psfrag}
\usepackage{slashed}

\def\la{\langle}

\definecolor{mygreen}{rgb}{0,0.7,0}

\def\spAB#1#2#3{\la#1|#2|#3]}

\DeclareMathOperator{\tr}{\rm tr}

\def\MP#1#2{(#1\cdot#2)}

\def\fmNf{\left( 4-n_f\right)}
\def\omNfpNs{\left( 1-n_f+n_s\right)}
\def\tmNs{\left( 3-n_s\right)}

\def\eps{\epsilon}
\def\sg(#1){\textrm{sign}(#1)}

\def\fl#1{#1^\flat}

\def\kf#1{K_{#1}^\flat}
\def\qb{\bar{q}}

\def\MHVb{$\overline{\rm MHV}$}
\def\dbox{\text{dbox}}
\def\xbox{\text{xbox}}
\def\pbox{\text{pbox}}

\def\vec#1{#1}
\def\w{\omega}

\def\fl#1{{#1^{\flat}}}
\def\flm#1{{#1^{\flat,\mu}}}
\def\kf#1{{\fl{K_{#1}}}}
\def\kfm#1{{\flm{K_{#1}}}}

\def\ulim#1{\underset{#1}{\lim}}

\newcommand{\id}{\mathrm{d}}

\newcommand{\nn}{\nonumber}

\newcommand{\feyn}[1]{#1\kern-0.45em/}
\newcommand{\A}[1]{\langle #1 \rangle}
\newcommand{\B}[1]{\left[ #1 \right]}

\newcommand{\fed}{\boldsymbol}

\preprint{}

\title{Hepta-Cuts of Two-Loop Scattering Amplitudes}

\author[a]{Simon Badger}
\author[a]{Hjalte Frellesvig}
\author[a]{Yang Zhang}
\affiliation[a]{
Niels Bohr International Academy and Discovery Center, The Niels Bohr Institute,\\%
University of Copenhagen, Blegdamsvej 17, DK-2100 Copenhagen, Denmark}
\emailAdd{badger@nbi.dk,hjf@nbi.dk,zhang@nbi.dk}

\abstract{
We present a method for the computation of hepta-cuts of two loop scattering amplitudes.
Four dimensional unitarity cuts are used to factorise the integrand onto the product of six tree-level
amplitudes evaluated at complex momentum values. Using Gram matrix constraints we derive a general parameterisation of the
integrand which can be computed using polynomial fitting techniques. The resulting expression is 
further reduced to master integrals using conventional integration by parts methods. 
We consider both planar and non-planar topologies for $2\to2$ scattering processes and apply the method to compute 
hepta-cut contributions to gluon-gluon scattering in Yang-Mills theory with adjoint fermions and scalars. 
}


\begin{document}

\maketitle
\flushbottom

\section{Introduction}

Precision cross section predictions for hadron colliders are an essential tool in the search for new
physics. While one-loop amplitudes give access to quantitative Next-to-Leading Order (NLO)
background estimates, a reliable analysis of the theoretical uncertainty requires
Next-to-Next-to-Leading Order (NNLO) corrections.

The use of Feynman diagram techniques for the evaluation of scattering amplitudes has always
presented a major challenge owing to the rapid growth in complexity with increasing loop order and
external legs. In recent years the development of on-shell methods
\cite{Bern:1994zx,Bern:1994cg,Cachazo:2004kj,Britto:2004nc,Britto:2004ap} has played a major role in removing this traditional
bottleneck for both tree-level and one-loop amplitudes.\footnote{See \cite{Ellis:2011cr,Ita:2011hi,Feng:2011gc} and
references therein for recent reviews on these topics.} Generalised unitarity
\cite{Bern:1995db,Britto:2004nc,Britto:2005ha,Britto:2006sj,Mastrolia:2006ki,Anastasiou:2006jv,Anastasiou:2006gt,Forde:2007mi,Ellis:2007br,Giele:2008ve,Badger:2008cm} and
integrand reduction techniques (OPP) \cite{Ossola:2006us} have been developed into fully automated
numerical algorithms able to compute high multiplicity one-loop amplitudes
\cite{Ossola:2007ax,Berger:2008sj,Giele:2008bc,Ellis:2008qc,Giele:2009ui,Mastrolia:2010nb,Badger:2010nx,Hirschi:2011pa,Bevilacqua:2011xh,Cullen:2011ac}.

Unitarity methods for multi-loop amplitudes have proven to be extremely powerful tools in
super-symmetric gauge theories. Maximal cutting techniques are an efficient method for reducing
these complicated amplitudes to the evaluation of a limited number of master integrals. These
techniques have been developed in the course of gluon-gluon scattering amplitudes in $\mathcal{N}=4$
Super-Yang-Mills (SYM) enabling computations up to five loops
\cite{Bern:1997nh,Bern:2005iz,Bern:2006ew,Bern:2007ct}.\footnote{We note that only the coefficients
of the master integrals are known at five loops, not the integrals themselves.} At two loops
computations with up to six external legs have been achieved
\cite{Bern:2008ap,Kosower:2010yk,Dixon:2011nj}.
Octa-cuts \cite{Buchbinder:2005wp} and the related leading singularity method
\cite{Cachazo:2008vp,Cachazo:2008hp} are also valid approaches in $\mathcal{N}=4$ SYM enabling the
computation of loop amplitudes directly from tree-level input.

In non super-symmetric theories, like QCD, the basis of integrals is far more complicated yet the
current state-of-the art techniques have been able to compute $2\to2$ processes in massless QCD 
\cite{Bern:1997nh,Bern:2000dn,Anastasiou:2000kg,Anastasiou:2000ue,Anastasiou:2001sv,Glover:2001af,Bern:2002tk}.
The motivation for the present study is to use some of the technology successful in the
super-symmetric cases to simplify the computation of these amplitudes.

The aim is to construct full two-loop amplitudes from products of tree-level amplitudes following
the successful approach taken at one-loop. Following a top down approach one begins with the
leading singularities, then systematically reduces the number of cuts to study more of the full amplitude.
At each step one subtracts the singularity structure previously constructed in order to obtain a
polynomial system. At one-loop this procedure relies on the knowledge of a basis of
integral functions. Though such a basis is not known at two loops, the reduction of arbitrary loop integrals
can be understood using integration by parts (IBP) relations
\cite{Tkachov:1981wb,Chetyrkin:1981qh}. Using a restricted set of IBPs constructed using Gr\"obner
bases, Gluza, Kajda and Kosower were able to construct a unitarity compatible integral basis for
planar topologies \cite{Gluza:2010ws}. Schabinger recently showed similar sets of IBPs could be obtained without the
use of Gr\"obner bases \cite{Schabinger:2011dz}. To date both a maximal unitarity approach
\cite{Kosower:2011ty} and an integrand reduction program similar to OPP have been proposed
\cite{Mastrolia:2011pr} which explore the use of fitting such a basis from tree-level input.

The approach we follow here will allow us to construct a general integrand parameterisation from
analysis of Gram matrices. This system can be matched to an expansion of the products of tree-level
amplitudes evaluated at a complete set of on-shell solutions to the loop momenta. This leads to a
linear system of equations that can be inverted to derive a master formula for the reduction of the
integrand. The part of the integrand that remains after integration is compatible with further reduction
to master integrals by using conventional IBP identities and Lorentz Invariance identities \cite{Gehrmann:1999as} 
by means of the Laporta algorithm \cite{Laporta:2001dd}, or
the related approaches \cite{Smirnov:2005ky,Smirnov:2006tz,Lee:2008tj}. A number of public tools
\cite{Anastasiou:2004vj,Smirnov:2008iw,Studerus:2009ye,vonManteuffel:2012yz} are available to
perform this step of the computation.

We address the first in a long list of ingredients required for a general decomposition of two loop amplitudes,
the maximum singularities in $2\to2$ processes. These are all contributions with seven propagators that can be 
extracted from hepta-cuts in four dimensions. The procedure reduces the
computation of the amplitude to a polynomial fitting procedure over a product of tree-level
amplitudes and is amenable to both analytic and numerical techniques. We compare with
the super-symmetric results obtained using the recent approach of Kosower and Larsen \cite{Kosower:2011ty}.

Our paper is organised as follows. We begin by re-deriving some results of the generalised unitarity
algorithm at one-loop. We focus on some of the key issues that we will apply to the two-loop case.
We then turn out attention to the three independent seven propagator topologies for $2\to2$
processes, the planar double box and penta-box configurations as well as the non-planar crossed box
configuration.  We develop a method for determination of an integrand parameterisation
using constraints from $5\times5$ Gram matrices. We then use this information to construct an
invertible linear system from the full set of on-shell solutions which maps the this
parameterisation to products of tree-level amplitudes.  The resulting integrand can be further
reduced to master integrals by application of well known integration by parts identities.  We
demonstrate the technique by applying it to the four-gluon scattering in Yang-Mills theory. The
expressions can be related to those in super-symmetric Yang-Mills and we comment on some the
simplifications that occur in those cases.  Finally we present our conclusions and some outlook for
future studies.

\section{Review of Generalised Unitarity at One-Loop}

In this section we will re-derive the well known integrand parameterisation
used in numerical one-loop generalised unitarity algorithms \cite{Ellis:2007br} and the closely
related integrand reduction of Ossola, Papadopoulos and Pittau (OPP) \cite{Ossola:2006us}. For
more detailed reviews of the subject we refer the reader to refs. \cite{Ellis:2011cr,Ita:2011hi}.

We represent a general ordered one-loop amplitude as a product of rational coefficients
multiplying scalar integral functions with four or fewer propagators. For the present
exercise we will restrict ourselves to cases where all numerators are in four dimensions and
the pentagon contributions are collected into a remaining rational contribution,
\begin{align}
  A_n^{(1)} =& 
  \int \frac{\id^d \vec{k}}{(2 \pi)^{d}}
  \sum_{i_1=1}^{n-3}
  \sum_{i_2=i_1+1}^{n-2}
  \sum_{i_3=i_2+1}^{n-1}
  \sum_{i_4=i_3+1}^{n} 
  \frac{\Delta_{4,i_1i_2i_3i_4}(\vec{k})}{D_{i_1} D_{i_2} D_{i_3} D_{i_4} }
  \nonumber\\&
  +\sum_{i_1=1}^{n-2}
  \sum_{i_2=i_1+1}^{n-1}
  \sum_{i_3=i_2+1}^{n} 
    \frac{\Delta_{3,i_1i_2i_3}(\vec{k})}{D_{i_1} D_{i_2} D_{i_3} }
  +\sum_{i_1=1}^{n-2}
  \sum_{i_2=i_1+2}^{n+i_1-2} 
    \frac{\Delta_{2,i_1i_2}(\vec{k})}{D_{i_1} D_{i_2} }
  \nonumber\\&
  +\text{tadpoles, wave-function bubbles and rational terms}.
  \label{eq:oneloopintegrand}
\end{align}
In the above we have defined the inverse propagators $D_{i_x} = (\vec{k}-\vec{p}_{i_1,i_x-1})^2$ and the dimension
$d=4-2\eps$. We define $\vec{p}_{i,j} = \sum_{k=i}^j\vec{p}_k$ as the sum of external momenta such that $\vec{p}_{i_1,i_1-1}=0$ and
have taken the restriction that all propagators are massless.

A general one-loop amplitude can be computed by repeated evaluations of a process specific numerator 
once general forms for the integrands, $\Delta_{c,X}(\vec{k})$, have been constructed. The numerator
used to fit the cut integrands could be generated from a Feynman diagram representation but in the
following we will take a top down approach and factorise each cut into products of tree-level amplitudes.

We will go through this known procedure in some detail since our generalisation to two loops will
follow it closely.

\subsection{Quadruple Cuts \label{sec:1lD4}}

\begin{figure}[h]
  \begin{center}
    \psfrag{1}{$\vec{P}_1$}
    \psfrag{2}{$\vec{P}_2$}
    \psfrag{3}{$\vec{P}_3$}
    \psfrag{4}{$\vec{P}_4$}
    \psfrag{l1}{$\vec{k}$}
    \psfrag{l2}{}
    \psfrag{l3}{}
    \psfrag{l4}{}
    \includegraphics[width=5cm]{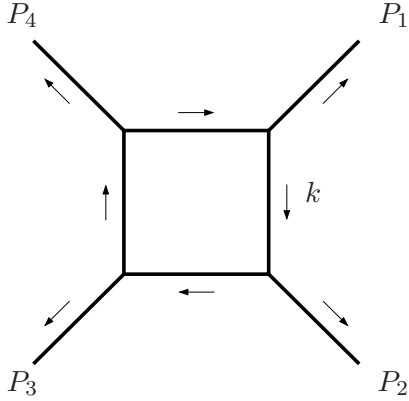}
  \end{center}
  \caption{Conventions for the momentum flow in the one-loop box.}
  \label{fig:box}
\end{figure}

Quadruple cuts of the one-loop amplitudes were first considered in the work of Britto, Cachazo and
Feng \cite{Britto:2004nc}. Each integrand, $\Delta_{4,i_1i_2i_3i_4}$, depends upon three independent
external momenta, say $\{\vec{P}_1=p_{i_4,i_1-1},\vec{P}_2=p_{i_1,i_2-1},\vec{P}_3=p_{i_2,i_3-1}\}$
as indicated in fig. \ref{fig:box}. 

In order to span the full four dimensional space of the integrand we are able to define a vector $\vec{\w}$, 
satisfying $\vec{\w}\cdot\vec{P}_k=0$ and $\w^2>0$. Such a direction can be called \textit{spurious} since,
\begin{equation}
  \int \frac{\id^d \vec{k}}{(2 \pi)^{d}} \frac{k\cdot\w}{D_{i_1}D_{i_2}D_{i_3}D_{i_4}} = 0.
  \label{eq:spuriousint}
\end{equation}
A simple representation for this complex vector is given by the totally anti-symmetric
tensor $\w^\mu\propto\varepsilon^{\mu123}$. Equivalently, using the basis of massless vectors ($\kf1,\kf2$) constructed
from $P_1$ and $P_2$ as described in Appendix \ref{app:spinors}, we can write:
\begin{equation}
  \w^\mu(P_3) = \frac{1}{2\gamma_{12}}\left(\spAB{\kf2}{P_3}{\kf1}\spAB{\kf1}{\gamma^\mu}{\kf2} - \spAB{\kf1}{P_3}{\kf2}\spAB{\kf2}{\gamma^\mu}{\kf1}\right).
  \label{eq:D4spuriousvector}
\end{equation}
This gives a set of scalar products with which we can write
down a completely general form of the cut integrand,
\begin{equation}
  \{(\vec{k}\cdot \vec{k}),(\vec{k}\cdot \vec{P}_1),
  (\vec{k}\cdot \vec{P}_2),
  (\vec{k}\cdot \vec{P}_3),
  (\vec{k}\cdot \vec{\w})\}.
\end{equation}
We are able to re-write some of the dot products in terms of inverse propagators and constant factors,
\begin{align}
  (\vec{k}\cdot \vec{k}) &= D_{i_1}, \\
  2 (\vec{k}\cdot \vec{P}_1) &= D_{i_4} -  D_{i_1} - P_1^2, \\
  2 (\vec{k}\cdot \vec{P}_2) &= D_{i_1} -  D_{i_2} + P_2^2, \\
  2 (\vec{k}\cdot \vec{P}_3) &= D_{i_2} -  D_{i_3} + 2 P_2\cdot P_3 + P_3^2,
  \label{eq:toprop}
\end{align}
where all $D_{i_x}$ vanish when the on-shell conditions are applied,
\begin{equation}
  \{\vec{k}^2=0,(\vec{k}-\vec{P}_2)^2=0,(\vec{k}-\vec{P}_2-\vec{P}_3)^2=0,(\vec{k}+\vec{P}_1)^2=0\}.
  \label{eq:1lbox-onshell}
\end{equation}
This leaves us with one irreducible scalar product (ISP),
\begin{equation}
  \Delta_{4,i_1i_2i_3i_4}(\vec{k}) = \sum_{\alpha} c_{\alpha}
  (\vec{k}\cdot \vec{\w})^{\alpha}.
  \label{eq:1lD4}
\end{equation}
Renormalizability tells us that $\alpha<4$, yet we must find another relation before we are in a
position to apply the cuts. Such information can be simply extracted from Gram matrices
\cite{Ellis:2007br}. For $2n$ vectors $\{l_1, \ldots, l_n; v_1,\ldots, v_n\}$, the $n\times n$ Gram matrix $G$ is defined as
\begin{equation}
  \label{eq:1}
  G\equiv G\left(
    \begin{array}{ccc}
      l_1 &\ldots& l_n\\
      v_1 & \ldots & v_n
   \end{array}
\right) ,\quad G_{ij}=l_i \cdot v_j.
\end{equation}
In particular, for the case where
$\{l_1,\ldots, l_n\}$ is identical to $\{ v_1, \ldots, v_n\}$, we define
\begin{equation}
  \label{eq:4}
  G(l_1, \ldots, l_n)\equiv G\left(\begin{array}{ccc}
l_1,\ldots, l_n\\
l_1,\ldots, l_n
\end{array}\right).
\end{equation}
The determinant $\det G$ is linear and anti-symmetric in the vectors in each row,
\begin{eqnarray}
  \label{eq:2}
 && \det G\left(
    \begin{array}{ccc}
      l_1+l_1' &\ldots& l_n\\
      v_1 & \ldots & v_n
   \end{array}
\right)= \det G\left(
    \begin{array}{ccc}
      l_1 &\ldots& l_n\\
      v_1 & \ldots & v_n
   \end{array}
\right)+\det G\left(
    \begin{array}{ccc}
      l_1' &\ldots& l_n\\
      v_1 & \ldots & v_n
   \end{array}
\right),
\label{gram-linear}
\\
&&\det G\left(
    \begin{array}{cccc}
      l_1 & l_2 & \ldots& l_n\\
      v_1 & v_2 & \ldots & v_n
   \end{array}
\right)=-\det G\left(
    \begin{array}{cccc}
      l_2 & l_1 & \ldots& l_n\\
      v_1 & v_2 & \ldots & v_n
   \end{array}.
\label{gram-antisymmetric}
\right)
\end{eqnarray}
Therefore $\det G$ vanishes if $\{l_1, \ldots, l_n\}$ or $\{ v_1, \ldots, v_n\}$ are linearly dependent. 

A Gram matrix can be used to calculate the inner products of two
vectors, if their projection on a given basis is known. Let
$\{e_1, ..., e_d\}$ span a $d$-dimensional vector space and let $l$ and
$v$ be two vectors in that space. Once we expand $v=v_i e_i$ and
$l=l_i e_i$ and define $G_d$=$G(e_1,..., e_d)$ we can write,
\begin{eqnarray}
  \label{eq:3}
  \left(
    \begin{array}{c}
      v\cdot e_1 \\
         \vdots\\
      v \cdot e_d
    \end{array}
\right)=G_d  \left(
    \begin{array}{c}
      v_1 \\
         \vdots\\
      v_d
    \end{array}
\right),
     \end{eqnarray}
and so,
\begin{equation}
 (l \cdot v)=(l_1, \ldots, l_d) G_d \left(
    \begin{array}{c}
      v_1 \\
         \vdots\\
      v_d
    \end{array}
\right)=(l\cdot e_1, \ldots, l\cdot e_d) G_d^{-1} \left(
    \begin{array}{c}
      v\cdot e_1 \\
         \vdots\\
      v\cdot e_d
    \end{array}
\right).
\label{quadratic-gram}
\end{equation}
If the basis \(\vec{e}_i\) is orthonormal, the Gram matrix becomes the identity and the relation
eq.\eqref{quadratic-gram} is trivial.  

Explicitly, let $d=4$ and $\{P_1,P_2,P_3,\omega \}$ be the basis of
4-dimensional loop momenta. When $k$ is $4$-dimensional momenta, by eq.\eqref{quadratic-gram},
we obtain a quadratic relation of the
Lorentz invariants,
\begin{eqnarray}
  k^2&=&(k \cdot P_1, k \cdot P_2, k \cdot P_3, k \cdot
  \omega)G_4^{-1} (k \cdot P_1, k \cdot P_2, k \cdot P_3, k \cdot
  \omega)^T   \label{quadratic-gram-kk-1}.
\end{eqnarray}
Alternatively, when $k$ is a $4$-dimensional loop momenta, by
the linear dependence property,
\begin{eqnarray}
 \det G\left(
    \begin{array}{ccccc}
      P_1 &P_2 &P_3 &\omega &k\\
      P_1 &P_2 &P_3 &\omega &k
    \end{array}
\right)=0.
\label{quadratic-gram5-kk-1}
\end{eqnarray}
It is easy to see that (\ref{quadratic-gram5-kk-1}),
is equivalent
to (\ref{quadratic-gram-kk-1}). 
%This equation is going to be the starting point for the relations
%between the invariant scalar products.In this quadruple cut example
%we choose $\vec{v}=\{\vec{P}_1,\vec{P}_2,\vec{P}_3,\vec{\w}\}$ 
We can get a non-trivial relation by computing the on-shell constraint $k^2=0$,
\begin{equation}
  (\vec{k}\cdot \vec{\w})^2 =-\w^2 (k \cdot P_1, k \cdot P_2, k \cdot P_3)G_3^{-1} (k \cdot P_1, k \cdot P_2, k \cdot P_3)^T=\text{const.} 
\end{equation}
where $G_3=G(P_1,P_2,P_3)$ is a constant matrix and $(k\cdot P_i)$, $i=1,2,3$ are all constant at the quadruple cut. This relation tells us the $\alpha=0,1$ in eq. \eqref{eq:1lD4}.

%Such information can be simply extracted from Gram matrices
%\cite{Ellis:2007br} which are defined by,
%\begin{align}
%G_{ij} & \define \vec{v}_i \cdot \vec{v}_j
%\label{eq:gramdef}
%\end{align}
%It has the property that the projection of some vector \(\vec{k} = \sum_i l_i \vec{v}_i\) into this
%basis can be written as
%\begin{align}
%\left( 
%\begin{array}{c} 
 % \vec{k} \cdot \vec{v}_1 \\
  %\vdots \\ 
  %\vec{k} \cdot \vec{v}_d 
%\end{array} 
%\right)
%&= G 
%\left( 
%\begin{array}{c} 
 % \vec{k}_1 \\
  %\vdots \\
  %\vec{k}_d 
%\end{array} 
 % \right)
%\label{eq:dotprel}
%\end{align}
%If the basis \(\vec{v}_i\) is orthonormal, the Gram matrix becomes the identity and the relation
%eq.\eqref{eq:dotprel} is trivial.  The dot product between two vectors can be written \(\vec{l}
%\cdot \vec{l}' = l_i G_{ij} l_j'\) which combined with the inverse of eq.\eqref{eq:dotprel} gives
%\begin{align}
 % \vec{l} \cdot \vec{l}' &= \left( \vec{l} \cdot \vec{v}_1 \cdots \vec{l} \cdot \vec{v}_d \right) G^{-1} 
%\left( 
%\begin{array}{c} 
 % \vec{l}'_1\cdot \vec{v}_1 \\
  %\vdots \\
  %\vec{l}'_d\cdot \vec{v}_n 
%\end{array} 
%\right) 
%\label{eq:GramRelation}
%\end{align}
%This equation is going to be the starting point for the relations between the invariant scalar products.
%In this quadruple cut example we choose $\vec{v}=\{\vec{P}_1,\vec{P}_2,\vec{P}_3,\vec{\w}\}$ so we can get 
%a non-trivial relation by considering computing the on-shell constraint $l^2=0$ via eq. \eqref{eq:GramRelation}:
%\begin{equation}
 % (\vec{k}\cdot \vec{\w})^2 = \text{const.},
%\end{equation}
%which tells us the $\alpha=0,1$ in eq.\eqref{eq:1lD4}.

This is the maximum number of constraints available for this topology so we can turn our attention to the on-shell solutions
for the loop momentum which will allow us to fit the coefficients $c_0$ and $c_1$, the only coefficients left in \eqref{eq:1lD4} by the constraints.

Following the well known parameterisation from the literature in terms of two-component Weyl spinors
we find two complex solutions for $\vec{k}$ satisfying the constraints of eq.
\eqref{eq:1lbox-onshell}.
On each solutions, of which explicit forms are given in Appendix \ref{app:1lbox}, the amplitude
factorises onto a product of tree amplitudes, $\mathcal{T}_{i_1i_2i_3i_4}$,
\begin{align}
  \Delta_{i_1i_2i_3i_4}(\vec{k}^{(s)}) = c_0 + c_1(\vec{k}^{(s)}\cdot\vec{\w})  = \mathcal{T}_{i_1i_2i_3i_4}(\vec{k}^{(s)}) \equiv d_s.
  \label{eq:box-ddef}
\end{align}
where,
\begin{align}
  \mathcal{T}_{x_1\cdots x_n}(\vec{k}) = 
  \sum_{\lambda_k=\pm}
  \prod_{k=1}^{n} 
  A^{(0)}\left( 
  \left( -\vec{k}+\vec{P}_{x_1,x_k-1} \right)^{-\lambda_k},
  \vec{p}_{x_k},\ldots,\vec{p}_{x_{k+1}-1},
  \left( \vec{k}-\vec{P}_{x_1,x_{k+1}-1} \right)^{\lambda_{k+1}}
  \right),
  \label{eq:1ltreeproduct}
\end{align}
with $n+1\equiv1$. Since the four on-shell constraints, eq. \eqref{eq:1lbox-onshell}, freeze the loop momentum, $d_s$ are just complex numbers.
From the explicit solutions (see Appendix \ref{app:1lbox}) we find $(\vec{k}^{(1)}\cdot\vec{\w})=\sqrt{V_4}$ 
and $(\vec{k}^{(2)}\cdot\vec{\w})=-\sqrt{V_4}$ which, after feeding into eq. \eqref{eq:box-ddef}, leads quickly to the final result:
\begin{align}
  c_0 &= \frac{1}{2}\left( d_1+d_2 \right), \\
  c_1 &= \frac{1}{\sqrt{V_4}}\left( d_1-d_2 \right).
\end{align}
The important feature of this analysis was the use of the Gram matrix to constrain the form of the
integrand which was then mapped to the products of tree amplitudes via the on-shell solutions to the
loop momentum. For this simple case the number of loop momentum solutions and the number of
independent coefficients in the integrand were the same. As we will see in the next example this
feature is not always true.

\subsection{Triple Cuts \label{sec:1ltri}}

To re-derive the formula for the triple cut integrands, $\Delta_{3,i_1i_2i_3}$, we follow exactly the same
procedure as above. This time our space is spanned by two external momenta and two trivial space
vectors $v=\{\vec{P}_1=p_{i_3,i_1-1},\vec{P}_2=p_{i_1,i_2-1},\vec{\w}_1,\vec{\w}_2\}$ where in the $(\kf1,\kf2)$ basis we can write,
\begin{align}
  \w_1^\mu &= \frac{1}{2}\left(
   \spAB{\kf1}{\gamma^\mu}{\kf2}
  +\spAB{\kf2}{\gamma^\mu}{\kf1}
  \right), \\
  \w_2^\mu &= \frac{i}{2}\left(
   \spAB{\kf1}{\gamma^\mu}{\kf2}
  -\spAB{\kf2}{\gamma^\mu}{\kf1}
  \right).
  \label{eq:D3spuriousvectors}
\end{align}
Removing the trivial scalar products that can be re-written in terms of propagators we have the
following form for the integrand,
\begin{equation}
  \Delta_{3,i_1i_2i_3}(\vec{k}) = \sum_{\alpha,\beta} c_{\alpha\beta} (\vec{k}\cdot \vec{\w}_1)^\alpha(\vec{k}\cdot \vec{\w}_2)^\beta.
\end{equation}
Renormalizability implies that $\alpha+\beta \leq 3$ and therefore there are ten
$c_{ij}$ coefficients. The Gram matrix identity, 
\begin{eqnarray}
  \det G\left(
  \begin{array}{ccccc}
    P_1 & P_2 & \w_2 &\w_2 &k\\
    P_1 & P_2 & \w_1 &\w_2 &k
  \end{array}
  \right)=0 ,
  \label{eq:TriGram}
\end{eqnarray}
leads us to the relation,
\begin{equation}
  (\vec{k}\cdot \vec{\w}_1)^2+(\vec{k}\cdot \vec{\w}_2)^2 = \text{const.}
\end{equation}
which reduces the number of independent $c_{\alpha\beta}$ coefficients to seven. In principle we could
chose any seven of the ten but in order to ensure terms in the integrand proportional to $(\vec{k}\cdot
\vec{\w}_k)$ integrate to zero, it is convenient to choose:
\begin{align}
  \bf c &= (c_{00},c_{01},c_{10},c_{20;02},c_{11},c_{12},c_{21}),
\end{align}
so that integrand is written,
\begin{align}
  \Delta_{3,i_1i_2i_3}(\vec{k}) &= 
  c_{00}
  +c_{10}(\vec{k}\cdot \vec{\w}_1) 
  +c_{01}(\vec{k}\cdot \vec{\w}_2)
  +c_{11}(\vec{k}\cdot \vec{\w}_1) (\vec{k}\cdot \vec{\w}_2)\nonumber\\&
  +c_{12}(\vec{k}\cdot \vec{\w}_1) (\vec{k}\cdot \vec{\w}_2)^2
  +c_{21}(\vec{k}\cdot \vec{\w}_1)^2 (\vec{k}\cdot \vec{\w}_2)\nonumber\\&
  +c_{20;02}\left( (\vec{k}\cdot \vec{\w}_1)^2 - (\vec{k}\cdot \vec{\w}_2)^2 \right).
  \label{eq:1lD3}
\end{align}
We then turn to the on-shell constraints,
\begin{equation}
  \{\vec{k}^2=0,\,(\vec{k}-\vec{P}_1)^2=0,\,(\vec{k}-\vec{P}_1-\vec{P}_2)^2=0\},
  \label{eq:1ltri-onshell}
\end{equation}
of which there are two possible solutions:
\begin{align}
  k^{(1),\mu} &= 
  a_{1} \kfm1 + a_{2} \kfm2 + \frac{t}{2} \spAB{\kf1}{\gamma^\mu}{\kf2} 
  + \frac{a_{1}a_{2}}{2t} \spAB{\kf2}{\gamma^\mu}{\kf1}, \\
  k^{(2),\mu} &= 
  a_{1} \kfm1 + a_{2} \kfm2 + \frac{t}{2} \spAB{\kf2}{\gamma^\mu}{\kf1} 
  + \frac{a_{1}a_{2}}{2t} \spAB{\kf1}{\gamma^\mu}{\kf2},
  \label{eq:1lD3-loopmom}
\end{align}
where,
\begin{align}
  a_{1} =  \frac{P_2^2\left( \gamma_{12}+P_1^2 \right)}{\gamma_{12}^2-P_1^2P_2^2} \; , \qquad a_{2} = - \frac{P_1^2\left( \gamma_{12}+P_2^2 \right)}{\gamma_{12}^2-P_1^2P_2^2}.
  \label{eq:1lD3-loopcoeffs}
\end{align}
By feeding this into eq.\eqref{eq:1lD3} we define the coefficients $d_{s,x}$ which
can be extracted from the subtracted product of three tree-level amplitudes,
\begin{equation}
  \Delta_{3,i_1i_2i_3}(\vec{k}^{(s)}) = \Delta^{(s)}_{3,i_1i_2i_3}(t) = \sum_{x=-3}^3 d_{s,x} t^x.
\end{equation}
Equating coefficients of $t$ on both sides of this equation gives us a $14\times7$ matrix, $M$,
which relates the $d_{s,x}$ coefficients to the $c_{k}$ coefficients,
\begin{equation}
  {\bf d} = M\cdot {\bf c}
\end{equation}
where ${\bf d} =
(d_{1,-3},d_{1,-2},d_{1,-1},d_{1,0},d_{1,1},d_{1,2},d_{1,3},d_{2,-3},d_{2,-2},d_{2,-1},d_{2,0},d_{2,1},d_{2,2},d_{2,3})$.
The final step is to invert $M$, the fact that it is invertible means that we have a unitarity cut
compatible parameterisation of the integrand. The inverse falls into two regions, firstly
when all $P_k^2\neq0$ (corresponding to $a_{1}a_{2}\neq0$) the null space of $M$ contains all of the $k^{(2)}$ solution and
$c$ coefficients are,
\begin{align}
  c_{00} &= d_{1,0} \\
  c_{10} &= 
      d_{1,1} 
      - a_{1}a_{2} d_{1,3} 
      - \frac{1}{a_{1}a_{2}} d_{1,-1} 
      - \frac{1}{a_{1}^2a_{2}^2} d_{1,-3}
  \\
  c_{01} &= 
      i d_{1,1} 
      + i a_{1}a_{2} d_{1,3} 
      + \frac{i}{a_{1}a_{2}} d_{1,-1} 
      + \frac{i}{a_{1}^2a_{2}^2} d_{1,-3}
      \\
  c_{11} &= 2id_{1,2}+\frac{2i}{a_{1}^2a_{2}^2} d_{1,-2} \\
  c_{20;02} &= d_{1,2}-\frac{1}{a_{1}^2a_{2}^2} d_{1,-2} \\
  c_{21} &= 4id_{1,3}+\frac{4i}{a_{1}^3a_{2}^3} d_{1,-3} \\
  c_{12} &= -4d_{1,3}+\frac{4}{a_{1}^3a_{2}^3} d_{1,-3}.
\end{align}
In case any $P_k^2=0$ we will find $a_{1}a_{2}=0$ and all $d$'s with negative powers are zero and the only non trivial equation
in the null space of $M$ is that $d_{2,0}=d_{1,0}$, the $c$ coefficients are,
\begin{align}
  c_{00} &= d_{1,0} \\
  c_{10} &= d_{1,1}-d_{2,1}\\
  c_{01} &= i\left(d_{1,1}+d_{2,1}\right) \\
  c_{11} &= 2i\left( d_{1,2}+d_{2,2} \right)\\
  c_{20;02} &= d_{1,2}-d_{2,2}\\
  c_{21} &=  4i\left( d_{1,3}+d_{2,3} \right)\\
  c_{12} &=  -d_{1,3}+d_{2,3}
\end{align}
Of course at this stage we in full agreement with the known results from Refs.
\cite{Ellis:2007br,Ossola:2006us}. 

As a final remark we consider the possibility of fitting the integrand using the large momentum
limit, $t\to\infty$, matching to the method of Forde \cite{Forde:2007mi}. This derivation follows the
argument presented in the recent review article of Ellis, Kunszt, Melnikov and Zanderighi \cite{Ellis:2011cr}. 
The subtraction terms, coming from box contributions previously evaluated with quadruple cuts, can be 
written schematically as,
\begin{equation}
  \mathcal{S}(\vec{k}) = \sum_j \frac{\Delta_{4,i_1i_2i_3 j}(\vec{k})}{D_j},
\end{equation}
where $\Delta_{4,i_1i_2i_3 j}(\vec{k})$ is given in eq.\eqref{eq:1lD4} using $\w^\mu(P_j)$ in
eq.\eqref{eq:D4spuriousvector}.  The integrand is formed from the two scalar products,
\begin{align}
  2\vec{k}^{(1)}\cdot \vec{\w}(P_j) &= \left(\spAB{\kf1}{P_j}{\kf2} t - \frac{a_{1}a_{2}}{t} \spAB{\kf2}{P_j}{\kf1} \right), \\
  2\vec{k}^{(2)}\cdot \vec{\w}(P_j) &= \left(-\spAB{\kf2}{P_j}{\kf1} t + \frac{a_{1}a_{2}}{t} \spAB{\kf1}{P_j}{\kf2} \right).
\end{align}
Explicitly taking the limit $t\to\infty$ yields,
\begin{equation}
  \ulim{t\to\infty} \mathcal{S}^{(s)}(t) = (-1)^s \frac{1}{2} c_{1;i_1i_2i_3j} + \mathcal{O}(t^{-1}).
\end{equation}
Since we know,
\begin{equation}
  \mathcal{T}_{i_1i_2i_3}(\vec{k})-\mathcal{S}^{(s)}(t) = \sum_{x=-3}^3 d_{s,x} t^x
\end{equation}
we can define a new set of coefficients $d'_{s,x}$
\begin{equation}
  \ulim{t\to\infty}\left(\mathcal{T}_{i_1i_2i_3}(\vec{k}^{(s)})\right) = \sum_{x=0}^3 d'_{s,x} t^x + \mathcal{O}(t^{-1}),
\end{equation}
such that $d'_{s,0}=d_{s,0}+\tfrac{(-1)^s}{2} c_{1;i_1i_2i_3j}$ and $d'_{s,x}=d_{s,x}$ for all $x>0$.
Inverting gives,
\begin{align}
  c_{00} = d_{1,0} = d_{2,0} = \frac{1}{2}\left( d'_{1,0}+d'_{2,0} \right).
\end{align}
enabling us to extract the non-spurious coefficient from the triple-cut alone as derived in refs.
\cite{Forde:2007mi,Ellis:2011cr}.

No further subtleties arise for double (or even single) cuts so
we won't reproduce any further results at one-loop.

\section{Integrand Representations of Two Loop Amplitudes}

The extension of the generalised unitarity algorithm from one to two loops is complicated by the
fact that a general loop integral basis is not known. However, it has be known for some time how to
reduce a general two-loop Feynman diagram to a basis of master (no longer simply scalar) integrals
by the use of integration by parts identities. For a cut based construction of the amplitude, such a
basis is unfortunately not suitable since the doubled and crossed propagators that appear do not
factorise onto simple poles, and hence neither onto the products of tree-level amplitudes.

A unitarity compatible basis has been explored using a Gr\"obner basis construction in a recent
paper by Gluza, Kajda and Kosower \cite{Gluza:2010ws}. Here will follow a slightly different
approach in which instead of trying to fit the coefficients of a minimal basis of functions, we will
fit an integrand level expression compatible with unitarity cuts, which can be further reduced to
master integrals by any of the known techniques. This is similar to the integrand reduction
technique recently presented by Mastrolia and Ossola \cite{Mastrolia:2011pr}.

For the purposes of this initial study we focus on the parts of the amplitude sensitive to seven 
propagator cuts in four dimensions. Though a small step towards a full integrand level
reduction technique we will emphasise some of the features that we hope apply to a wider class of
cuts.  For the present paper we will be concerned with primitive amplitudes contributing to
gluon-gluon scattering two loops. In particular this restricts us to the case where no subtraction terms from
octa-cuts is required though the procedure is expected to follow in a similar fashion
\cite{Mastrolia:2011pr}. The colour ordered partial amplitudes are defined by \cite{Bern:1997nh},
\begin{align}
  \mathcal{A}_4^{(2)} &= g_s^6 
  \sum_{\sigma\in S_4/Z_4} 
  \tr(
  T^{a_{\sigma(1)}}
  T^{a_{\sigma(2)}}
  T^{a_{\sigma(3)}}
  T^{a_{\sigma(4)}}
  )
  \Big( 
  N^2 A^{(2),LC}_{4;1,1}(\sigma(1),\sigma(2),\sigma(3),\sigma(4))
  \nonumber\\&
  \hspace{6.5cm}
  + A^{(2),SC}_{4;1,1}(\sigma(1),\sigma(2),\sigma(3),\sigma(4)) 
  \Big)
  \nonumber\\&
  + g_s^6\sum_{\sigma\in S_4/Z_4^3} 
  N_c\tr(
  T^{a_{\sigma(1)}}
  T^{a_{\sigma(2)}}
  )
  \tr(
  T^{a_{\sigma(3)}}
  T^{a_{\sigma(4)}}
  )
  A^{(2)}_{4;1,3}(\sigma(1),\sigma(2);\sigma(3),\sigma(4))
\end{align}
where $N_c$ is the number of colours and $T^{a_i}$ are the fundamental generators of $SU(N_c)$.
These partial amplitudes are mapped to primitive amplitudes before unitarity cuts can be applied
using colour ordered tree-level amplitudes.
\begin{align}
  &A^{(2),LC}_{4;1,1}(1,2,3,4) = 
    A_4^{[\dbox]}(1,2;3,4;)
  + A_4^{[\dbox]}(2,3;4,1;)
  + A_4^{[\pbox]}(1;2,3,4;) \nonumber\\&
  + A_4^{[\pbox]}(2;3,4,1;)
  + A_4^{[\pbox]}(3;4,1,2;)
  + A_4^{[\pbox]}(4;1,2,3;)
  \\%
  &A^{(2),SC}_{4;1,1}(1,2,3,4) =
  2 A_4^{[\dbox]}(1,2;3,4;)
  + 2 A_4^{[\dbox]}(1,2;4,3;)
  - 4 A_4^{[\dbox]}(1,3;2,4;) \nonumber\\&
  + 2 A_4^{[\dbox]}(2,3;4,1;)
  - 4 A_4^{[\dbox]}(2,4;3,1;)
  + 2 A_4^{[\dbox]}(3,2;4,1;) \nonumber\\&
  + 2 A_4^{[\xbox]}(3;2,1;4)
  - 4 A_4^{[\xbox]}(2;3,1;4)
  + 2 A_4^{[\xbox]}(2;4,1;3) \nonumber\\&
  + 2 A_4^{[\xbox]}(1;2,3;4)
  - 4 A_4^{[\xbox]}(1;2,4;3)
  + 2 A_4^{[\xbox]}(1;3,4;2)
  \\%
  &A^{(2)}_{4;1,3}(1,2;3,4) =
  6 A_4^{[\dbox]}(1,2;3,4;)         
  + 6 A_4^{[\dbox]}(1,2;4,3;)         
  + 2 A_4^{[\pbox]}(1;2,3,4;) \nonumber\\&         
  + 2 A_4^{[\pbox]}(1;3,4,2;)         
  + 2 A_4^{[\pbox]}(1;4,2,3;)         
  + 2 A_4^{[\pbox]}(2;1,3,4;) \nonumber\\&         
  + 2 A_4^{[\pbox]}(2;3,4,1;)         
  + 2 A_4^{[\pbox]}(2;4,1,3;)         
  + 2 A_4^{[\pbox]}(3;1,2,4;) \nonumber\\&
  + 2 A_4^{[\pbox]}(3;2,4,1;)
  + 2 A_4^{[\pbox]}(3;4,1,2;)
  + 2 A_4^{[\pbox]}(4;1,2,3;) \nonumber\\&
  + 2 A_4^{[\pbox]}(4;2,3,1;)
  + 2 A_4^{[\pbox]}(4;3,1,2;)
  + 4 A_4^{[\xbox]}(3;2,1;4) \nonumber\\&
  - 2 A_4^{[\xbox]}(2;3,1;4)
  - 2 A_4^{[\xbox]}(2;4,1;3)
  - 2 A_4^{[\xbox]}(1;2,3;4) \nonumber\\&
  - 2 A_4^{[\xbox]}(1;2,4;3)
  + 4 A_4^{[\xbox]}(1;3,4;2)
  \label{eq:2lprimitivedecomp}
\end{align}
The decomposition involves three topologies: the double box, the crossed box and the penta-box which
we reference explicitly in the superscript for clarity. 
We will examine the four dimensional hepta-cut part of these primitives in the following section. 
We label each integrand with a subscript for the number of cut propagators and a set of indices
labelling the momenta leaving the diagram at each vertex. A '$*$' label indicates that no external
momentum enters a vertex. Using this notation, which is described in more detail in appendix
\ref{app:notation}, the primitive amplitudes are written:
\begin{align}
  A_4^{[\dbox]}(1,2;3,4;) &= 
  \int\int \frac{\id^d \vec{k}}{(2 \pi)^{d}} \frac{\id^d \vec{q}}{(2 \pi)^{d}}
  \Bigg(
  \frac{\Delta^{\dbox}_{7;12*34*}(\vec{k},\vec{q})}{\prod_{k=1}^{7} l_k^2}
  \Bigg)+\ldots\\
  A_4^{[\pbox]}(1;2,3,4;) &= 
  \int\int \frac{\id^d \vec{k}}{(2 \pi)^{d}} \frac{\id^d \vec{q}}{(2 \pi)^{d}}
  \Bigg(
  \frac{\Delta^{\pbox}_{7;1*234*}(\vec{k},\vec{q})}{\prod_{k=1}^{7} l_k^2}
  \Bigg)+\ldots\\
  A_4^{[\xbox]}(1;3,4;2) &= 
  \int\int \frac{\id^d \vec{k}}{(2 \pi)^{d}} \frac{\id^d \vec{q}}{(2 \pi)^{d}}
  \Bigg(
  \frac{\Delta^{\xbox}_{7;1*34*2}(\vec{k},\vec{q})}{\prod_{k=1}^{7} l_k^2}
  \Bigg)+\ldots
  \label{eq:2lamp}
\end{align}
where `$\ldots$' represents terms with $\leq$ 6 propagators and terms only accessible via
$d$-dimensional cuts. The above decomposition only applies to the pure gluonic loops but we may also
use it for gluino and adjoint scalar loops. 

\section{Hepta-cuts of Two-Loop Amplitudes}

\begin{figure}[h]
  \begin{center}
    \psfrag{1}{$1$}
    \psfrag{2}{$2$}
    \psfrag{3}{$3$}
    \psfrag{4}{$4$}
    \psfrag{l1}{$l_2$}
    \psfrag{l2}{$l_3$}
    \psfrag{l3}{$l_4$}
    \psfrag{l4}{$l_5$}
    \psfrag{l5}{$l_6$}
    \psfrag{l6}{$l_1$}
    \psfrag{l7}{$l_7$}
    \includegraphics[width=6.5cm]{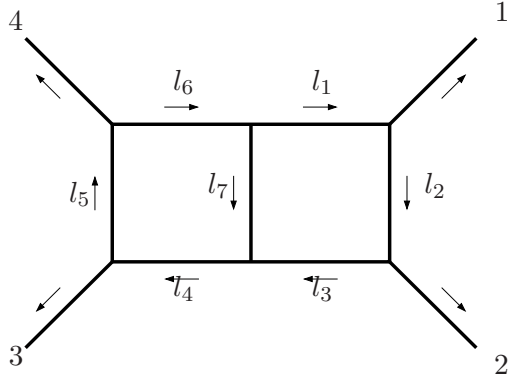}
  \end{center}
  \caption{Conventions for the momentum flow in the planar double box.}
  \label{fig:dblbox}
\end{figure}

We will proceed with the integrand reduction through a three step process which utilises:
\begin{itemize}
  \item Relations from Gram matrices to find a general form for the integrand.
  \item Finding the total number of on-shell solutions, these will be families of solutions
    depending on a number of free parameters.
  \item After fitting the full integrand, further reduction of non-spurious terms to Master
    Integrals (MIs) can be achieved using IBP relations
\end{itemize}
The full integrand can then be constructed as the solution to a linear system of equations. In the
following we go through the details of the three independent seven propagator topologies for four-point amplitudes with
massless legs: the double box (shown in fig. \ref{fig:dblbox}), the crossed box (shown in fig.
\ref{fig:xbox}) and the penta-box (shown in fig. \ref{fig:pbox}).

\subsection{Integrand Parameterisations from Gram Matrix Constraints}

Gram matrices are also important for two loop amplitude computation. Let $k,
q$ be the loop momenta and $\{e_1,e_2,e_3,e_4\}$ be the basis of
4-dimensional momenta. When $k$ and $q$ are $4$-dimensional momenta,
we obtain three quadratic relations of the
Lorentz invariants using eq. \eqref{quadratic-gram},
\begin{eqnarray}
  k^2&=&(k \cdot e_1, k \cdot e_2, k \cdot e_3, k \cdot
  e_4)G_4^{-1} (k \cdot e_1, k \cdot e_2, k \cdot e_3, k \cdot
  e_4)^T   \label{quadratic-gram-11}\\
 q^2&=&(q \cdot e_1, q \cdot e_2, q \cdot e_3, q \cdot
  e_4)G_4^{-1} (q \cdot e_1, q \cdot e_2, q \cdot e_3, q \cdot
  e_4)^T\label{quadratic-gram-22}\\
 p\cdot q&=&(k \cdot e_1, k \cdot e_2, k \cdot e_3, k \cdot
  e_4)G_4^{-1} (q \cdot e_1, q \cdot e_2, q \cdot e_3, q\cdot
  e_4)^T\label{quadratic-gram-12}
\end{eqnarray}
Alternatively, when $k$ and $q$ are $4$-dimensional-momenta, by
the linear dependence property,
\begin{eqnarray}
 \lambda_{kk}\equiv \det G\left(
    \begin{array}{ccccc}
      e_1 &e_2 &e_3 &e_4 &k\\
      e_1 &e_2 &e_3 &e_4 &k
    \end{array}
\right)=0  \label{quadratic-gram5-11}\\
  \lambda_{qq}\equiv \det G\left(
    \begin{array}{ccccc}
     e_ 1 &e_2 &e_3 &e_4 &q\\
      e_1 &e_2 &e_3 &e_4 &q
    \end{array}
\right)=0  \label{quadratic-gram5-22}\\
 \lambda_{kq}\equiv 
 \det G\left(
    \begin{array}{ccccc}
     e_1 &e_2 &e_3 &e_4 &k\\
     e_1 &e_2 &e_3 &e_4 &q
    \end{array}
\right)=0  \label{quadratic-gram5-12}
\end{eqnarray}
It is easy to see that (\ref{quadratic-gram5-11}),
(\ref{quadratic-gram5-22}) and (\ref{quadratic-gram5-12}) are equivalent
to (\ref{quadratic-gram-11}), (\ref{quadratic-gram-22}) and
(\ref{quadratic-gram-12}), respectively. 

It seems that there are many more $5\times 5$ Gram matrix relations. Because we
can choose $5$ vectors from the set $\{e_1,e_2,e_3,e_4,k,q\}$
twice, there are 
\begin{equation}
  \tfrac{1}{2}\Big(
  \tbinom{6}{5}+1
  \Big)
  \tbinom{6}{5}
  =21
\end{equation}
$5\times 5$ Gram-matrix relations for $4$-dimensional loop momenta. However, the additional
relations are not independent, since they can be generated by  (\ref{quadratic-gram5-11}),
(\ref{quadratic-gram5-22}) and (\ref{quadratic-gram5-12}), by the linear and anti-symmetric
properties of Gram matrices, (\ref{gram-linear}) and (\ref{gram-antisymmetric}).

To see this, we consider $k$ and $q$ as general $d$-dimensional
vectors,
\begin{eqnarray}
  \label{eq:6}
  k=\sum_{i=1}^4 k_{i} e_i+ k^\mathbf{n}, \quad q=\sum_{i=1}^4 q_{i} e_i+ q^\mathbf{n},
\end{eqnarray}
where $k^\mathbf{n}$ and $q^\mathbf{n}$ are the extra-dimensional
components. It is clear that
\begin{eqnarray}
  \label{eq:8}
  \lambda_{kk}&=&\det G\left(
    \begin{array}{ccccc}
      e_1 &e_2 &e_3 &e_4 &k^\mathbf{n}\\
      e_1 &e_2 &e_3 &e_4 &k^\mathbf{n}
    \end{array}
\right)=\det(G_4)  (k^\mathbf{n})^2.\\
  \lambda_{qq}&=&\det G\left(
    \begin{array}{ccccc}
      e_1 &e_2 &e_3 &e_4 &q^\mathbf{n}\\
      e_1 &e_2 &e_3 &e_4 &q^\mathbf{n}
    \end{array}
\right)=\det(G_4)  (q^\mathbf{n})^2.\\
 \lambda_{kq}&=&\det G\left(
    \begin{array}{ccccc}
      e_1 &e_2 &e_3 &e_4 &k^\mathbf{n}\\
      e_1 &e_2 &e_3 &e_4 &q^\mathbf{n}
    \end{array}
\right)=\det(G_4)  (k^\mathbf{n} \cdot q^\mathbf{n}).
\end{eqnarray}

For other $5\times 5$ Gram matrices, comparing with the above expressions,
\begin{eqnarray}i
 \det G\left(
    \begin{array}{ccccc}
      e_1 &e_2 &e_3 &k &q\\
      e_1 &e_2 &e_3 &e_4 &q
    \end{array}
\right)&=&  k_{4} \det G\left(
    \begin{array}{ccccc}
      e_1 &e_2 &e_3 & e_4 &q^\mathbf{n}\\
      e_1 &e_2 &e_3 &e_4 &k^\mathbf{n}
    \end{array}
\right)+q_{4} \det G\left(
    \begin{array}{ccccc}
      e_1 &e_2 &e_3 & k^\mathbf{n} & e_4\\
      e_1 &e_2 &e_3 &e_4 &k^\mathbf{n}
    \end{array}
\right)\nonumber \\&+& \det G\left(
    \begin{array}{ccccc}
      e_1 &e_2 &e_3 & k^\mathbf{n} & q^\mathbf{n}\\
      e_1 &e_2 &e_3 &e_4 &k^\mathbf{n} 
    \end{array}\right)= k_{4} \lambda_{kq}-q_{4} \lambda_{kk},
\\
\det G\left(
    \begin{array}{ccccc}
      e_1 &e_2 &e_3 &k &q\\
      e_1 &e_2 &e_3 &k &q
    \end{array}
\right)&=& q_{4}^2 \lambda_{kk}+k_{4}^2 \lambda_{qq}-2 k_{4} q_{4}
\lambda_{kq}+\frac{\det(G_3)}{\det(G_4)^2} (\lambda_{kk}\lambda_{qq}-\lambda_{kq}^2),
\end{eqnarray}
with other Gram matrices following a similar pattern. So as long as the relations (\ref{quadratic-gram5-11}),
(\ref{quadratic-gram5-22}) and (\ref{quadratic-gram5-12}) hold, all
the other $5\times 5$ Gram matrices will vanish automatically. However, in
practice, we will still use other $5\times 5$ Gram matrix
relations, since they usually provide very efficient ways of combining
the three 
fundamental relations (\ref{quadratic-gram5-11}),
(\ref{quadratic-gram5-22}) and (\ref{quadratic-gram5-12}) together.
\subsection{The Double Box}

We define the double box contribution to $A_4^{\text{planar},(2)}$ by the following propagators:
\begin{align}
  l_1^{\dbox} &= k & l_2^{\dbox} &= k-p_1 & l_3^{\dbox} &= k-p_{1,2} & l_4^{\dbox} &= -q+p_{3,4}  \nonumber\\
  l_5^{\dbox} &= -q+p_4 & l_6^{\dbox} &= -q & l_7^{\dbox} &= -q-k &&
  \label{eq:dbox-props}
\end{align}
Just as in the one-loop box topology, the three external momenta $\{\vec{p}_1,\vec{p}_2,\vec{p}_4\}$
are supplemented by the spurious vector, $\vec{\w}$, given in eq. \eqref{eq:D4spuriousvector}. The loop momentum is then contained in the space
spanned by $\vec{v}= \{ \vec{p}_1,\vec{p}_2,\vec{p}_4,\vec{\w} \}$. Taking into account scalar products that
can be trivially rewritten in terms of the propagators $(l_k^{\dbox})^2$ using relations of the form of
eq.\eqref{eq:toprop}, we can parameterise the integrand,
$\Delta^{\dbox}_{7;12*34*}(\vec{k},\vec{q})$, with four irreducible scalar products (ISPs) combined
into terms of the form,
\begin{equation}
 (k \cdot p_4)^m (q \cdot p_1)^n (k \cdot \w)^\alpha (q \cdot \w)^\beta.
  \label{eq:dboxISPs}
\end{equation}
The first constraints on the indices $m,n,\alpha$ and $\beta$ come from renormalization conditions
implying $m+n+\alpha+\beta \leq 6$. Since each of the integrals involves four propagators for this
topology we may also deduce $m+\alpha\leq4$ and $n+\beta\leq4$.

The Gram matrix relations (\ref{quadratic-gram5-11}),
(\ref{quadratic-gram5-22}) and (\ref{quadratic-gram5-12}) put
constraints on the ISP's. For the basis
$\{ e_1,e_2,e_3,e_4 \} = \{ p_1,p_2,p_4,\omega \} $, at the hepta-cut (\ref{quadratic-gram5-11})
reads,
\begin{equation}
(k\cdot\omega)^2=(k\cdot p_4-s_{14}/2)^2 
\end{equation}
This relation requires that $\alpha=0,1$. Similar,  (\ref{quadratic-gram5-22})
reads,
\begin{equation}
(q\cdot\omega)^2=(q\cdot p_1-s_{14}/2)^2
\end{equation}
So $\beta=0,1$. (\ref{quadratic-gram5-12}) requires that $\alpha
\beta=0$, since it reads,
\begin{equation}
(k\cdot\omega) (q\cdot\omega) =-\frac{s_{14}^2}{4} + \frac{s_{14} (k\cdot p_4)}{2} + \frac{s_{14} (q\cdot p_1)}{2} + \bigg(1+\frac{2s_{14}}{s_{12}}\bigg)  (k\cdot p_4)(q\cdot p_1)
\end{equation}
The number of the ISP monomials is reduced to $56$.

So far, we just used the three fundamental $5\times 5$ Gram-matrix
relation individually. It is possible to combine them together to get
more constraints. The efficient way is to consider other $5\times 5$ Gram-matrix
relations directly. We have, 
\begin{eqnarray}
  \det G\left(
    \begin{array}{ccccc}
      1 & 2 & 4 & k & q\\
 1 & 2 & 4 & k & q
    \end{array}
\right)=0,\quad
  \det G\left(
    \begin{array}{ccccc}
      1 & 2 & 4 & k & q\\
 1 & 2 & 4 & \omega & k
    \end{array}
\right)=0,\quad
  \det G\left(
    \begin{array}{ccccc}
      1 & 2 & 4 & k & q\\
 1 & 2 & 4 & \omega & q
    \end{array}
\right)=0
\label{Gram5-dependent-dbox}
\end{eqnarray}
For example, the first equation in (\ref{Gram5-dependent-dbox}) explicitly
reads,
\begin{equation}
  \label{eq:11}
0=4 (k \cdot p_4)^2 (q \cdot p_1)^2+2 s_{12} (k \cdot p_4)^2 (q \
\cdot p_1)+2 s_{12} (k \cdot p_4) (q \cdot p_1)^2-s_{12} s_{14} (k \cdot p_4) \
(q \cdot p_1) 
\end{equation}
so the terms with both $m\geq 2$ and $n\geq 2$ are reduced. These relations further reduce the number of ISP monomials to $32$.
With all our constraints complete we arrive at a general parameterisation for the
double box integrand,
\begin{align}
   \Delta^{\dbox}_{7;12*34*}(\vec{k},\vec{q}) &= 
   \sum_{mn \alpha \beta} c_{m n (\alpha+2\beta)} (k \cdot p_4)^m (q \cdot p_1)^n (k \cdot \w)^\alpha (q \cdot \w)^\beta.
\label{eq:D7dboxA}
\end{align}
There are 16 non-spurious terms, i.e. those not proportional to
$(k\cdot\w)$ or $(q\cdot\w)$,
\begin{equation}
  (c_{000},c_{010},c_{100},c_{020},c_{110},c_{200},c_{030},c_{120},c_{210},c_{300},c_{040},c_{130},c_{310},c_{400},c_{140},c_{410})
  \label{eq:dbox_ns}
\end{equation}
and 16 spurious terms,
\begin{equation}
  (
  c_{001},c_{011},c_{101},c_{111},c_{201},c_{211},c_{301},c_{311},
  c_{002},c_{012},c_{102},c_{022},c_{112},c_{032},c_{122},c_{132}
  )
  \label{eq:dbox_s}
\end{equation}
The terms can be represented in form of a table:
\begin{center}
  \begin{tabular}[h]{c|ccccc}
    \multicolumn{6}{c}{$\alpha=0, \; \beta=0$} \\
    & $n$=0 & $n$=1 & $n$=2 & $n$=3 & $n$=4 \\
    \hline
    $m$=0 & \checkmark & \checkmark & \checkmark & \checkmark & \checkmark \\
    $m$=1 & \checkmark & \checkmark & \checkmark & \checkmark & \checkmark \\
    $m$=2 & \checkmark & \checkmark \\
    $m$=3 & \checkmark & \checkmark \\
    $m$=4 & \checkmark & \checkmark \\
  \end{tabular}
\end{center}
and for the spurious terms,
\begin{center}
  \begin{tabular}[h]{c|cccc}
    \multicolumn{5}{c}{$\alpha=1, \; \beta=0$} \\
    & $n$=0 & $n$=1 & $n$=2 & $n$=3 \\
    \hline
    $m$=0 & \checkmark & \checkmark \\
    $m$=1 & \checkmark & \checkmark \\
    $m$=2 & \checkmark & \checkmark \\
    $m$=3 & \checkmark & \checkmark \\
  \end{tabular}
  \begin{tabular}[h]{c|cccc}
    \multicolumn{5}{c}{$\alpha=0, \; \beta=1$} \\
    & $n$=0 & $n$=1 & $n$=2 & $n$=3 \\
    \hline
    $m$=0 & \checkmark & \checkmark & \checkmark & \checkmark \\
    $m$=1 & \checkmark & \checkmark & \checkmark & \checkmark \\
    $m$=2 \\
    $m$=3 \\
  \end{tabular}
\end{center}
Our next task is to use the full set of on-shell solutions to find a map to these coefficients from the products of
tree-level amplitudes

\subsubsection{Solutions to the on-shell constraints}

The solutions to the on-shell constraints $(l^{\dbox}_k)^2=0$ have been considered in
Refs.\cite{Kosower:2011ty,Mastrolia:2011pr}.  The six solutions can be parameterised using the same
two component Weyl spinor basis as used at one-loop:
\begin{align}
  l_2^\mu &= x_1 p_1^\mu + x_2 p_2^\mu + x_3 \frac{\langle p_1 | \gamma^{\mu} | p_2 ]}{2} + x_4 \frac{\langle p_2 | \gamma^{\mu} | p_1 ]}{2} \nn \\
  l_5^\mu &= y_1 p_3^\mu + y_2 p_4^\mu + y_3 \frac{\langle p_3 | \gamma^{\mu} | p_4 ]}{2} + y_4 \frac{\langle p_4 | \gamma^{\mu} | p_3 ]}{2}
\label{eq:dboxmom1}
\end{align}
With eight unknowns and seven equations, each of the solutions depends on a free parameter which we 
will denote as $\tau$. The choice of this parameter has been made to ensure that the integrand takes a simple polynomial
form.
\begin{align}
\begin{array}{|c|c|c|c|c|c|c|c|c|}
\hline
\text{Solution} & x_1 & x_2 & x_3 & x_4 & y_1 & y_2 & y_3 & y_4 \\ 
\hline
1 & 0 & 0 & \tfrac{\A{23}}{\A{13}} & 0 & 0 & 0 & \frac{\A{14}}{\A{13}}(1-\tau) & 0 \\
2 & 0 & 0 & 0 & \tfrac{\B{23}}{\B{13}} & 0 & 0 & 0 & \frac{\B{14}}{\B{13}}(1-\tau) \\
3 & 0 & 0 & \frac{\B{14}}{\B{24}}(\tau-1) & 0 & 0 & 0 & -\tfrac{\B{23}}{\B{24}} & 0 \\
4 & 0 & 0 & 0 & \frac{\A{14}}{\A{24}}(\tau-1) & 0 & 0 & 0 & -\tfrac{\A{23}}{\A{24}} \\
5 & 0 & 0 & 0 & \tfrac{\B{13}\B{24}\tau + \B{12}\B{34}}{\B{14}\B{13}\tau} & 0 & 0 & -\tfrac{\B{13}}{\B{14}}(1+\tau) & 0 \\
6 & 0 & 0 & \tfrac{\A{13}\A{24}\tau + \A{12}\A{34}}{\A{14}\A{13}\tau} & 0 & 0 & 0 & 0 & -\tfrac{\A{13}}{\A{14}}(1+\tau) \\
\hline
\end{array} \nn
\label{eq:dbox-ossol}
\end{align}
It is straightforward to feed these solutions in the general integrand expression $\Delta_{7;12*34*}(k,q)$ and
define a set of coefficients that can be extracted from the product of six tree level amplitudes,
\begin{equation}
  \Delta_{7;12*34*}(k^{(s)},q^{(s)}) = \Delta^{(s)}_{7;12*34*}(\tau) =
  \begin{cases}
    \sum_{x=0}^{4} d_{s,x} \tau^x & 1,2,3,4, \\
    \sum_{x=-4}^{4} d_{s,x} \tau^x & 5,6.
  \end{cases}
  \label{eq:dcoeffs}
\end{equation}
where,
\begin{align}
  \Delta_{7;12*34*}(q,k) &=
  \sum_{\lambda_k=\pm}
  A^{(0)}(-l_1^{-\lambda_1},p_1,l_2^{\lambda_2})
  A^{(0)}(-l_2^{-\lambda_2},p_2,l_3^{\lambda_3})
  A^{(0)}(-l_4^{-\lambda_4},p_3,l_5^{\lambda_5})\nonumber\\&\times
  A^{(0)}(-l_5^{-\lambda_5},p_4,l_6^{\lambda_6})
  A^{(0)}(-l_6^{-\lambda_6},l_1^{\lambda_1},l_7^{\lambda_7})
  A^{(0)}(-l_3^{-\lambda_3},l_4^{\lambda_4},-l_7^{-\lambda_7}).
\end{align}

We now follow the procedure used in section \ref{sec:1ltri} by constructing a $38\times32$ matrix
such that,
\begin{equation}
  \vec{\fed d} = \vec{M} \cdot \vec{\fed c}
\end{equation}
It is easy to check that this matrix has rank 32 and therefore a unique solution. We are able to invert the system using standard linear
algebra packages available for symbolic computations. The final list of equations mapping $d_{s,x}$
to $c_{mn(\alpha+2\beta)}$ is available in a computer readable format from {\tt http://www.nbia.dk/badger.html}. They have relatively
simple forms for example:
\begin{equation}
  c_{000} = \frac{1}{2}\left( d_{1,0} + d_{2,0} \right).
  \label{eq:dboxc000}
\end{equation}

\subsubsection{Integration by parts identities \label{sec:IBPdbox}}

Having obtained a method to fix the 32 coefficients of $\Delta_{7;12*34*}(k,q)$, it is now in a form
that can be further reduced to master integrals using integration by parts identities.
There are by now a number of packages available to perform the task of reducing the master
integrand, $\Delta_{7;12*34*}(k,q)$, onto a basis of two master integrals. For this purpose we
made use of the {\tt FIRE} Mathematica package \cite{Smirnov:2008iw}. In the case of the planar
double box this enables to compare our results directly with those of Kosower and Larsen
\cite{Kosower:2011ty}.
\begin{align}
  A_4^{[\dbox]}(1,2;3,4;) = 
  \int\int \frac{\id^d \vec{k}}{(2 \pi)^{d}} \frac{\id^d \vec{q}}{(2 \pi)^{d}}
  \frac{ C_1 + C_2 (k\cdot p_4) }{l^2_1l^2_2l^2_3l^2_4l^2_5l^2_6l^2_7} + \cdots
  \label{eq:2ldboxmasters}
\end{align}
where we suppress all further master integrals. In terms of our non spurious $c_{nm0}$ coefficients
they are,
\begin{align}
  C_1 &= c_{000} 
  + \frac{s_{12}s_{14}}{8}c_{110}
  - \frac{s_{12}^2s_{14}}{16}\left( c_{120}+c_{210} \right)
  + \frac{s_{12}^3s_{14}}{32}\left( c_{130}+c_{310} \right) \nonumber\\&
  - \frac{s_{12}^4s_{14}}{64}\left( c_{140}+c_{410} \right)
  , \\
  C_2 &= c_{100} + c_{010}
  - \frac{3s_{12}}{4}c_{110}
  + \frac{s_{14}}{2}\left( c_{020}+c_{200} \right)
  + \frac{3s_{12}^2}{8}\left( c_{120}+c_{210} \right) \nonumber\\&
  + \frac{s_{14}^2}{4}\left( c_{030}+c_{300} \right)
  - \frac{3s_{12}^3}{16}\left( c_{130}+c_{310} \right)
  + \frac{s_{14}^3}{8}\left( c_{040}+c_{400} \right) \nonumber\\&
  + \frac{3s_{12}^4}{32}\left( c_{140}+c_{410} \right)
  \label{eq:dbox-masterintcoeffs}
\end{align}
Closed form expressions for the master integrals can be found in refs.
\cite{Smirnov:1999gc,Smirnov:1999wz}.

\subsection{The Crossed Box}

\begin{figure}[h]
  \begin{center}
    \psfrag{1}{$1$}
    \psfrag{2}{$2$}
    \psfrag{3}{$3$}
    \psfrag{4}{$4$}
    \psfrag{l1}{$l_2$}
    \psfrag{l2}{$l_7$}
    \psfrag{l3}{$l_3$}
    \psfrag{l4}{$l_4$}
    \psfrag{l5}{$l_5$}
    \psfrag{l6}{$l_6$}
    \psfrag{l7}{$l_1$}
    \includegraphics[width=6.5cm]{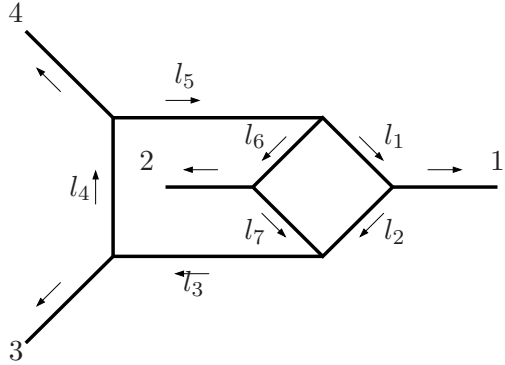}
  \end{center}
  \caption{Conventions for the momentum flow in the non-planar crossed box.}
  \label{fig:xbox}
\end{figure}

We represent crossed box topology shown in fig. \ref{fig:xbox} using the propagators as follows,
\begin{align}
  l_1^{\xbox} &= k+p_1 & l_2^{\xbox} &= k & l_3^{\xbox} &= q+p_3 & l_4^{\xbox} &= q  \nonumber\\
  l_5^{\xbox} &= q-p_4 & l_6^{\xbox} &= q-k+p_{2,3} & l_7^{\xbox} &= q-k+p_3 &&
  \label{eq:xbox-props}
\end{align}
As for the double box there are four ISPs parameterising the general integrand, $\Delta_{7;1*34*2}(k,q)$,
the Gram matrix constraints however lead us to a slightly different form of the final integrand.
Again, (\ref{quadratic-gram5-11}), (\ref{quadratic-gram5-22}) and (\ref{quadratic-gram5-12}) confine
$(\alpha,\beta)=(0,0), (1,0), (0,1)$, respectively, giving,
\begin{align}
   \Delta^{\xbox}_{7;1*34*2}(\vec{k},\vec{q}) &= \sum_{mn \alpha \beta} c_{m n (\alpha+2\beta)} (k \cdot p_3)^m (q \cdot p_2)^n (k \cdot \w)^\alpha (q \cdot \w)^\beta.
\label{eq:D7xboxA}
\end{align}
We can combine the three fundamental relations
together or consider other $5\times 5$ Gram matrix constraints, to reduce the dependent terms.
These relations have different form comparing with the double-box case, since there are fewer
symmetries in the crossed-box diagram. This leads us to a representation with 19 non-spurious terms,
\begin{equation}
  (c_{000},c_{010},c_{100},c_{020},c_{110},c_{200},c_{030},c_{120},c_{210},c_{300},c_{040},c_{220},c_{310},c_{400},c_{050},c_{320},c_{410},c_{060},c_{420})
  \label{eq:xbox-ns}
\end{equation}
and 19 spurious terms,
\begin{equation}
  (c_{001},c_{011},c_{101},c_{111},c_{201},c_{211},c_{301},c_{311},
  c_{002},c_{012},c_{102},c_{022},c_{112},c_{032},c_{122},c_{042},c_{132},c_{052},c_{142})
  \label{eq:xbox-s}
\end{equation}
The terms can be represented in form of a table:
\begin{center}
  \begin{tabular}[h]{c|ccccccc}
    \multicolumn{8}{c}{$\alpha=0, \; \beta=0$} \\
    & $n$=0 & $n$=1 & $n$=2 & $n$=3 & $n$=4 & $n$=5 & $n$=6 \\
    \hline
    $m$=0 & \checkmark & \checkmark & \checkmark & \checkmark & \checkmark & \checkmark & \checkmark \\
    $m$=1 & \checkmark & \checkmark & \checkmark \\
    $m$=2 & \checkmark & \checkmark & \checkmark \\
    $m$=3 & \checkmark & \checkmark & \checkmark \\
    $m$=4 & \checkmark & \checkmark & \checkmark \\
  \end{tabular}
\end{center}
and for the spurious terms,
\begin{center}
  \begin{tabular}[h]{c|cccccc}
    \multicolumn{7}{c}{$\alpha=1, \; \beta=0$} \\
    & $n$=0 & $n$=1 & $n$=2 & $n$=3 & $n$=4 & $n$=5\\
    \hline
    $m$=0 & \checkmark & \checkmark \\
    $m$=1 & \checkmark & \checkmark \\
    $m$=2 & \checkmark & \checkmark \\
    $m$=3 & \checkmark & \checkmark \\
  \end{tabular}
  \begin{tabular}[h]{c|cccccc}
    \multicolumn{7}{c}{$\alpha=0, \; \beta=1$} \\
    & $n$=0 & $n$=1 & $n$=2 & $n$=3 & $n$=4 & $n$=5\\
    \hline
    $m$=0 & \checkmark & \checkmark & \checkmark & \checkmark & \checkmark & \checkmark \\
    $m$=1 & \checkmark & \checkmark & \checkmark & \checkmark & \checkmark \\
    $m$=2 \\
    $m$=3 \\
  \end{tabular}
\end{center}

\subsubsection{Solutions to the on-shell constraints}

The crossed box topology uses the same basis as the planar box,
\begin{align}
l_2^\mu &= x_1 p_1^\mu + x_2 p_2^\mu + x_3 \frac{\langle p_1 | \gamma^{\mu} | p_2 ]}{2} + x_4 \frac{\langle p_2 | \gamma^{\mu} | p_1 ]}{2} \nn \\
l_4^\mu &= y_1 p_3^\mu + y_2 p_4^\mu + y_3 \frac{\langle p_3 | \gamma^{\mu} | p_4 ]}{2} + y_4 \frac{\langle p_4 | \gamma^{\mu} | p_3 ]}{2}
\label{eq:xboxmom1}
\end{align}
which we then use to for solve $\{l_k^2\}=0$.

The result is 8 families of solutions, again parameterised by $\tau$. These can be summarised by:
\begin{align}
\begin{array}{|c|c|c|c|c|c|c|c|c|}
\hline
\text{Solution} & x_1 & x_2 & x_3 & x_4 & y_1 & y_2 & y_3 & y_4 \\ 
\hline
1 & \tfrac{s_{14}+\tau}{s_{12}} & 0 & \tfrac{\A{23}(s_{13}+\tau)}{\A{13}s_{12}} & 0 
  & 0 & 0 & \tfrac{\tau}{\A{32}\B{24}} & 0 \\
2 & \tfrac{s_{14}+\tau}{s_{12}} & 0 & 0 & \tfrac{\B{23}(s_{13}+\tau)}{\B{13}s_{12}}
  & 0 & 0 & 0 & \tfrac{\tau}{\B{32}\A{24}} \\
3 & 0 & 0 & \frac{\tau}{\A{13}\B{32}} & 0
  & 0 & 0 & -\frac{\B{23}}{\B{24}} & 0 \\
4 & 0 & 0 & 0 & \frac{\tau}{\B{13}\A{32}}
  & 0 & 0 & 0 & -\frac{\A{23}}{\A{24}} \\
5 & \frac{s_{14}+\tau}{s_{12}} & 0 & 0 & -\tfrac{\A{13}(s_{14}+\tau)}{\A{23}s_{12}}
& 0 &  0 & \tfrac{\tau}{\B{24}\A{32}} & 0 \\  
6 & \frac{s_{14}+\tau}{s_{12}} & 0 & -\tfrac{\B{13}(s_{14}+\tau)}{\B{23}s_{12}} & 0
& 0 & 0 & 0 & \tfrac{\tau}{\A{24}\B{32}} \\  
7 & -1 & 0 & 0 & \tfrac{s_{14}+\tau}{\B{13}\A{32}}
  & 0  & 0 & \tfrac{\B{32}}{\B{14}} & 0 \\
8 & -1 & 0 & \tfrac{s_{14}+\tau}{\A{13}\B{32}} & 0
& 0 & 0 & 0 & \tfrac{\A{32}}{\A{14}} \\
\hline
\end{array} \nn
\label{eq:xbox-ossol}
\end{align}
From which, thanks to the choice of $\tau$ in eq. \eqref{eq:xbox-ossol}, we define the polynomial form of the cut integrand,
\begin{equation}
  \Delta_{7;1*34*2}(k^{(s)},q^{(s)}) = \Delta^{(s)}_{7;1*34*2}(\tau) =
  \begin{cases}
    \sum_{x=0}^{6} d_{s,x} \tau^x & s=1,2,5,6, \\
    \sum_{x=0}^{4} d_{s,x} \tau^x & s=3,4,7,8.
  \end{cases}
  \label{eq:xbox-dcoeffs}
\end{equation}
where,
\begin{align}
  \Delta_{7;1*34*2}(q,k) &=
  \sum_{\lambda_k=\pm}
  A^{(0)}(-l_1^{-\lambda_1},p_1,l_2^{\lambda_3})
  A^{(0)}(-l_6^{-\lambda_6},p_2,l_7^{\lambda_7})
  A^{(0)}(-l_3^{-\lambda_2},p_3,l_4^{\lambda_4})\nonumber\\&\times
  A^{(0)}(-l_4^{-\lambda_4},p_4,l_5^{\lambda_5})
  A^{(0)}(-l_5^{-\lambda_5},l_1^{\lambda_1},l_6^{\lambda_6})
  A^{(0)}(-l_2^{-\lambda_2},l_3^{\lambda_3},-l_7^{-\lambda_7}).
\end{align}
As before this leads to an invertible matrix this time $48\times38$,
\begin{equation}
  \vec{\fed d} = \vec{M}\cdot\vec{ \fed c}
\end{equation}
The final equations for $c_{nm(\alpha+2\beta)}$ are of similar complexity to those
in the double box topology for example:
\begin{align}
  c_{000} = \frac{1}{2}\left( d_{1,0}+d_{2,0} \right).
  \label{eq:xbox-c000}
\end{align}
The complete set of relations can be obtained from {\tt http://www.nbia.dk/badger.html}.

\subsubsection{Integration by parts identities \label{sec:IBPxbox}}

The integration by parts identities generated using {\tt FIRE} reduce $\Delta_{7;1*34*2}$
onto two seven propagator master integrals,
\begin{align}
  A_4^{[\xbox]}(1;3,4;2) = 
  \int\int \frac{\id^d \vec{k}}{(2 \pi)^{d}} \frac{\id^d \vec{q}}{(2 \pi)^{d}}
  \frac{ C_1 + C_2 (k\cdot p_3) }{l^2_1l^2_2l^2_3l^2_4l^2_5l^2_6l^2_7} + \cdots
  \label{eq:2lxboxmasters}
\end{align}
where $C_1$ and $C_2$ are given by:
\begin{align}
  C_1 &= c_{000} 
  +\frac{1}{16}s_{14}s_{13}  (
   c_{200}
  - c_{110}
  + 2c_{020} 
  )          
  \nonumber\\&
  + \frac{1}{32}s_{14}s_{13} (s_{14}-s_{13})  (
   c_{300} 
  - c_{210} 
  + c_{120} 
  - 2c_{030} 
  )
  \nonumber\\&
  + \frac{1}{16^2}(3 (s_{14}-s_{13})^2 + s_{12}^2) s_{14}s_{13}  (
    c_{400}
  - c_{310}
  + c_{220}
  + 2c_{040}
  )
  \nonumber\\&
  + \frac{1}{16^2}((s_{14}-s_{13})^2 + s_{12}^2) s_{14}s_{13} (s_{14}-s_{13})  (
   c_{320}
  - c_{410}
  - 2c_{050}
  )
  \nonumber\\&
  + \frac{1}{16^3}(5 (s_{14}-s_{13})^4 + 10 s_{12}^2 (s_{14}-s_{13})^2 + s_{12}^4) s_{14}s_{13}  (
   c_{420}
  + 2c_{060}
  )
  \\
  C_2 &=
  c_{100}
  - 2 c_{010}
  + \frac{3}{8}(s_{14}-s_{13})  (
   c_{200}
  - c_{110}
  + 2c_{020}
  )
  \nonumber\\&
  + \frac{1}{16}(2 (s_{14}-s_{13})^2 + s_{12}^2)  (
   c_{300}
  - c_{210}
  + c_{120}
  - 2c_{030}
  )
  \nonumber\\&
  + \frac{2}{16^2}(5 (s_{14}-s_{13})^2 + 7 s_{12}^2) (s_{14}-s_{13})  (
   c_{400}
  - c_{310}
  + c_{220}
  + 2c_{040}
  )
  \nonumber\\&
  + \frac{1}{16^2}(3 (s_{14}-s_{13})^4 + 8 s_{12}^2 (s_{14}-s_{13})^2 + s_{12}^4)  (
   c_{320}
  - c_{410}
  - 2c_{050}
  )
  \nonumber\\&
  + \frac{2}{16^3}
  \left(7 (s_{14}-s_{13})^4 + 30 s_{12}^2 (s_{14}-s_{13})^2 + 11 s_{12}^4\right) (s_{14}-s_{13})  (
   c_{420}
  + 2c_{060}
  )
  \label{eq:xboxMIcoeffs}
\end{align}
The integrals themselves have been computed using Mellin-Barnes techniques in refs.
\cite{Tausk:1999vh,Anastasiou:2000mf}.

\subsection{The Penta-Box}

\begin{figure}[h]
  \begin{center}
    \psfrag{1}{$1$}
    \psfrag{2}{$2$}
    \psfrag{3}{$3$}
    \psfrag{4}{$4$}
    \psfrag{l1}{$l_1$}
    \psfrag{l2}{$l_2$}
    \psfrag{l3}{$l_3$}
    \psfrag{l4}{$l_4$}
    \psfrag{l5}{$l_5$}
    \psfrag{l6}{$l_6$}
    \psfrag{l7}{$l_7$}
    \includegraphics[width=6.5cm]{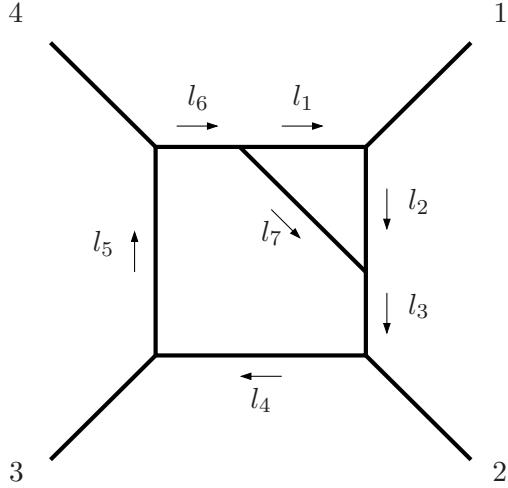}
  \end{center}
  \caption{Conventions for the momentum flow in the penta-box.}
  \label{fig:pbox}
\end{figure}

Our conventions for the penta-box topology follow those outlined in fig. \ref{fig:pbox}. The seven
propagators are,
\begin{align}
  l_1 &= q-k-p_4 & l_2&=q-k+p_{2,3} & l_3&= q+p_{2,3} & l_4 &= q+p_3 \nn \\
  l_5 &= q & l_6&=q-p_4 & l_7&=k
  \label{eq:pboxprops}
\end{align}
This topology has a rather different integrand structure than our previous cases. The Gram matrix relation,
\begin{equation}
  \label{eq:10}
  (k \cdot \w)= 2(q\cdot \w) \bigg(\frac{(k\cdot
    p_2)}{s_{12}}-\frac{(k\cdot p_4)}{s_{14}}\bigg)
\end{equation}
implies that $(k \cdot \w)$ is not independent at the hepta-cut. 
So we only have three ISPs, $(k\cdot p_2),(k\cdot p_4)$ and $(q\cdot\w)$, with a final form
parameterisation,
\begin{equation}
  \Delta^{\pbox}_{7;1*234*} = \sum_{m,n,\alpha} c_{mn\alpha} 
  (k\cdot p_2)^m (k\cdot p_4)^n(q\cdot\w)^\alpha.
  \label{eq:D7pbox}
\end{equation}
The sums are restricted such that there are 20 coefficients in all. Ten are non-spurious,
\begin{equation}
  \{c_{000},c_{100},c_{010},c_{020},c_{110},c_{200},c_{030},c_{120},c_{210},c_{300}\},
\end{equation}
and ten are spurious,
\begin{equation}
  \{c_{001},c_{101},c_{011},c_{021},c_{111},c_{201},c_{031},c_{121},c_{211},c_{301}\}.
\end{equation}
In the tabular format this looks like,
\begin{center}
  \begin{tabular}[h]{c|cccc}
    \multicolumn{5}{c}{$\alpha=0$} \\
    & $n$=0 & $n$=1 & $n$=2 & $n$=3 \\
    \hline
    $m$=0 & \checkmark & \checkmark & \checkmark & \checkmark \\
    $m$=1 & \checkmark & \checkmark & \checkmark  \\
    $m$=2 & \checkmark & \checkmark \\
    $m$=3 & \checkmark  \\
  \end{tabular}
  \begin{tabular}[h]{c|cccccc}
    \multicolumn{5}{c}{$\alpha=1$} \\
    & $n$=0 & $n$=1 & $n$=2 & $n$=3 \\
    \hline
    $m$=0 & \checkmark & \checkmark & \checkmark & \checkmark \\
    $m$=1 & \checkmark & \checkmark & \checkmark \\
    $m$=2 & \checkmark & \checkmark \\
    $m$=3 & \checkmark \\
  \end{tabular}
\end{center}

\subsubsection{Solutions to the on-shell constraints}

We parameterise the loop momenta according to,
\begin{align}
l_2^\mu &= x_1 p_1^\mu + x_2 p_2^\mu + x_3 \frac{\langle p_1 | \gamma^{\mu} | p_2 ]}{2} + x_4 \frac{\langle p_2 | \gamma^{\mu} | p_1 ]}{2} \nn \\
l_5^\mu &= y_1 p_3^\mu + y_2 p_4^\mu + y_3 \frac{\langle p_3 | \gamma^{\mu} | p_4 ]}{2} + y_4 \frac{\langle p_4 | \gamma^{\mu} | p_3 ]}{2}.
\label{eq:pboxmom1}
\end{align}
Interestingly in this case we find that the set of cut constraints is degenerate and we have two
independent solutions parameterised by $\tau_1$ and $\tau_2$,
\begin{align}
\begin{array}{|c|c|c|c|c|c|c|c|c|}
\hline
\text{Solution} & x_1 & x_2 & x_3 & x_4 & y_1 & y_2 & y_3 & y_4 \\ 
\hline
1 & \tfrac{\tau_1}{s_{12}s_{14}} & 0 & \tfrac{\A{23}\left( s_{12}s_{14}+\tau_1+\tau_2 \right)}{\A{13}s_{12}s_{14}} & 0 & 0 & 0 & -\tfrac{\B{23}}{\B{24}} & 0 \\
2 & \tfrac{\tau_1}{s_{12}s_{14}} & 0 & 0 & \tfrac{\B{23}\left( s_{12}s_{14}+\tau_1+\tau_2 \right)}{\B{13}s_{12}s_{14}} & 0 & 0 & 0 & -\tfrac{\A{23}}{\A{24}} \\
\hline
\end{array} \nn
\label{eq:pbox-ossol}
\end{align}
The choice of $\tau_1$ and $\tau_2$ has been made such that the integrand has a symmetric form for
the ISP's and we are able to write down a simple form for the inverted $20\times20$ system:
\begin{align}
  c_{mn\alpha} = (-1)^{m}\, 2^{n+m+\alpha-1}\, \frac{ s_{12}^n s_{14}^m }{s_{14}^\alpha} \big( d_{1,m,n} + (-1)^\alpha d_{2,m,n} \big)
\end{align}
where $d_{s,m,n}$ is the coefficient of \(\tau_1^m \tau_2^n\) for solution $s$.

\subsubsection{Integration by parts identities}

After the application of further reduction via IBP relations it turns out that all penta-box integrands are
reducible to six propagator master integrals or simpler topologies. Since those master integrals
will also have contributions from hexa-cut configurations a complete study of the form of these
reduction identities will be postponed to future studies. Nevertheless, for the purposes
of a complete integrand level reduction these terms play an essential role.

\section{Applications to Gluon-Gluon Scattering \label{sec:4gamps}}

In this section we apply our technique above to $gg\to gg$ scattering amplitudes. These amplitudes
have been known from some time \cite{Glover:2001af,Bern:2002tk} and present something of a benchmark test of our method.
We compute the amplitudes in Yang-Mills theory with an arbitrary number of massless
gluinos ($n_f$) and scalars ($n_s$) in the adjoint representation. By considering specific configurations of the number of
fermion and scalar flavours we are able to cross check our results against the simpler ones obtained
in super-symmetric Yang-Mills theories, using:
\begin{align}
  &\left(n_f=4, n_s=3\right) \rightarrow \mathcal{N}=4 \text{ SYM}, \\
  &\left(n_f=2, n_s=1\right) \rightarrow \mathcal{N}=2 \text{ SYM}, \\
  &\left(n_f=1, n_s=0\right) \rightarrow \mathcal{N}=1 \text{ SYM}.
  \label{eq:SUSYmap}
\end{align}
%\begin{align}
%  &\mathcal{N}=4 \text{ SYM} \; \rightarrow \; \left(n_f=4, n_s=3\right), \\
%  &\mathcal{N}=2 \text{ SYM} \; \rightarrow \; \left(n_f=2, n_s=1\right), \\
%  &\mathcal{N}=1 \text{ SYM} \; \rightarrow \; \left(n_f=1, n_s=0\right).
%  \label{eq:SUSYmap}
%\end{align}
Note that we consider scalars to be complex so there are two degrees of freedom for each scalar flavour.

\begin{figure}[h]
  \begin{center}
    \psfrag{1}{$1$}
    \psfrag{2}{$2$}
    \psfrag{3}{$3$}
    \psfrag{4}{$4$}
    \psfrag{nf}{$n_f$}
    \psfrag{ns}{$n_s$}
    \psfrag{nf}{$n_f$}
    \psfrag{nf*ns}{$n_f n_s$}
    \includegraphics[width=\textwidth]{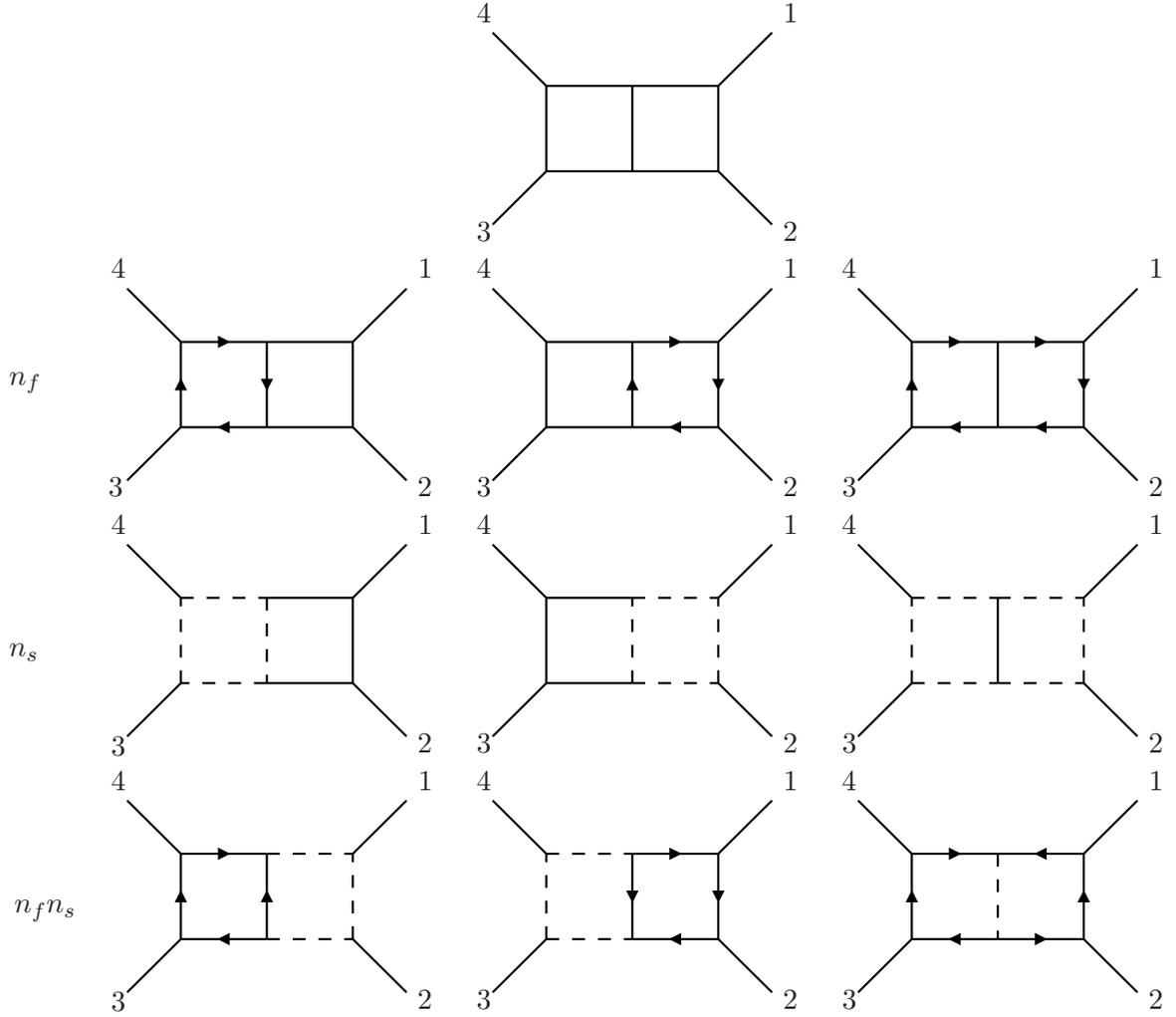}
  \end{center}
  \caption{Contributions to the double box in Yang-Mills theories of gluons, $n_f$ adjoint fermions and $n_s$
  scalars. Gluons are represented as solid lines, fermions as solid lines with arrows and scalars as
  dashed lines.}
  \label{fig:dboxSYM}
\end{figure}

Throughout this section we will use
\begin{equation}
  \Delta^{T}_{7;i_1\ldots i_6}(k,q) = 
  (\Delta_{7;i_1\ldots i_6}^{T, {\rm ns}}+\Delta_{7;i_1\ldots i_6}^{T, {\rm s}}) A^{(0)},
\end{equation}
where $\Delta_{7;i_1\ldots i_6}^{T, {\rm s}}$ contain all spurious dot products, which vanish after integration, 
and $\Delta_{7;i_1\ldots i_6}^{T, {\rm ns}}$ contains the remaining non-spurious dot products. The
four-point tree amplitudes $A^{(0)}$ are a shorthand for the functions $A^{(0)}(1^{\lambda_1},2^{\lambda_2},3^{\lambda_3},4^{\lambda_4})$
given in Appendix \ref{app:trees}. There are ten configurations of particles flowing in the loops
which contribute to the various topologies, the explicit cases of the double box are shown in fig.
\ref{fig:dboxSYM}.

The final results are factorised into terms that vanish depending on the amount of super-symmetry.
We note that in $\mathcal{N}=4$ the factors $\fmNf,\tmNs,\omNfpNs$ all vanish whereas $\omNfpNs$
will only appear in theories with no super-symmetry. Formulae as a function of the number of
super-symmetries, $\mathcal{N}$, can be obtained by:
\begin{align}
  n_f = \mathcal{N} \; , \qquad n_s = \mathcal{N}-1.
\end{align}

For the planar and non-planar double boxes we have cross checked the final coefficients of the
master integrals are in full agreement with the results by Bern, De Freitas and Dixon \cite{Bern:2002tk}.

\subsection{Planar Double Boxes}

The following section contain the results for the planar double box as defined in eq. \eqref{eq:D7dboxA}.

\subsubsection{The $--++$ Helicity Amplitude}

We find the following analytic forms for the integrands:
\begin{align}
  &\Delta_{7;12*34*}^{\dbox, {\rm ns},--++} = -s_{12}^2s_{14}
  \\
  &\Delta_{7;12*34*}^{\dbox, {\rm s},--++} = 0 
\end{align}
After applying further IBP relations as given in eq. \eqref{eq:dbox-masterintcoeffs},
the coefficients of the two basis integrals \cite{Gluza:2010ws,Kosower:2011ty} become:
\begin{align}
  C_1^{--++} &= -s_{12}^2s_{14} A^{(0)} \\
  C_2^{--++} &= 0
\end{align}

\subsubsection{The $-+-+$ Helicity Amplitude}

The integrands for this configuration are:
\begin{align}
  &\Delta_{7;12*34*}^{\dbox, {\rm ns},-+-+} = -s_{14} s_{12}{}^2 \nonumber\\&
  -\frac{\fmNf \tmNs s_{14} s_{12}{}^2 
  }{s_{13}{}^3}
  (\MP{k}{4}+\MP{q}{1}) \left(2 \MP{k}{4} \left(2 \MP{q}{1}+s_{12}\right)+s_{12} \left(2
  \MP{q}{1}-s_{14}\right)\right)
  \nonumber\\&
  +\frac{s_{14} s_{12}{}^2 \omNfpNs 
  }{s_{13}{}^4}
  \Big(
  16 \MP{k}{4}^3 \left(2\MP{q}{1}+s_{14}\right)
  +4 \MP{k}{4}^2 \left(6 s_{12} \MP{q}{1}+\left(2 s_{12}-s_{14}\right) s_{14}\right) \nonumber\\&
  +4 \MP{k}{4} \left(6 s_{12}\MP{q}{1}^2+s_{12} \left(2 s_{12}-s_{14}\right) \MP{q}{1}+8
  \MP{q}{1}^3-s_{12} s_{14}{}^2\right) \nonumber\\&
  -16 \MP{k}{4}^4-\left(s_{14} \left(2\MP{q}{1}+s_{12}\right)-4 \MP{q}{1}^2\right){}^2
  \Big)
  \nonumber\\&
  +\frac{\fmNf s_{14} s_{12}{}^2 
  }{2 s_{13}{}^3}
  \Big(
  4 \MP{k}{4}^2 \left(4\MP{q}{1}+3 s_{12}+s_{14}\right)
  +2 \MP{k}{4} \left(2 \MP{q}{1}-s_{14}\right) \left(4 \MP{q}{1}+3 s_{12}+s_{14}\right) \nonumber\\&
  +4 \left(3 s_{12}+s_{14}\right) \MP{q}{1}^2
  -2 s_{14} \left(3 s_{12}+s_{14}\right) \MP{q}{1}
  +s_{12} s_{13} s_{14}
  \Big)
\end{align}
\begin{align}
  &\Delta_{7;12*34*}^{\dbox, {\rm s},-+-+} = \nonumber\\&
  \frac{2 \fmNf \tmNs s_{14} s_{12}{}^2 
  }{s_{13}{}^3}
  \left(\MP{q}{1} \left(2 \MP{k}{4}+s_{12}\right) \MP{q}{\w}-\MP{k}{4} \MP{k}{\w}
  \left(2 \MP{q}{1}+s_{12}\right)\right)
  \nonumber\\&
  +\frac{s_{14} s_{12}{}^2 \omNfpNs 
  }{s_{13}{}^4}
  \Big( \nonumber\\&
  \MP{k}{\w} \big( 
  8 \MP{k}{4}^2 \left(4\MP{q}{1}+s_{14}\right)+8 \MP{k}{4} \left(3 s_{12} \MP{q}{1}+s_{14} \left(2 \MP{q}{1}+s_{12}\right)\right)
  \nonumber\\&
  -16 \MP{k}{4}^3+s_{12}{}^2
  \left(2 \MP{q}{1}+s_{14}\right)
  \big) \nonumber\\&
  +\MP{q}{\w} \big(
  -2 \MP{k}{4} \left(4 \left(3 s_{12}+2 s_{14}\right) \MP{q}{1}+16\MP{q}{1}^2+s_{12}{}^2\right)
  \nonumber\\&
  -s_{14} \left(8 s_{12} \MP{q}{1}+8 \MP{q}{1}^2+s_{12}{}^2\right)+16
  \MP{q}{1}^3
  \big)
  \Big)
  \nonumber\\&
  +\frac{\fmNf s_{12}{}^2 
  }{2 s_{13}{}^3}
  \Big(
  s_{12} s_{13} 
  (-\MP{k}{\w}) \left(2 \MP{q}{1}+s_{14}\right)
  \nonumber\\&
  +2 \MP{k}{4} \left(2 s_{14} \MP{k}{\w} \left(4 \MP{q}{1}+3 s_{12}+s_{14}\right)+\left(s_{12} s_{13}-8 s_{14} \MP{q}{1}\right)
  \MP{q}{\w}\right)
  \nonumber\\&
  +s_{14} \left(s_{12} s_{13}-4 \left(3 s_{12}+s_{14}\right) \MP{q}{1}\right) \MP{q}{\w}
  \Big)
\end{align}
which lead to coefficients of the master integrals of,

\begin{align}
  C_1^{-+-+} &=
  \frac{1}{4} s_{12}{}^2 s_{14} A^{(0)} 
  \left(
  -\frac{6 s_{12}{}^2 s_{14}{}^2 \omNfpNs}{s_{13}{}^4}
  +\frac{3 \fmNf s_{12} s_{14}}{s_{13}{}^2}
  -4
  \right)
  \\
  C_2^{-+-+} &=
  \frac{3 s_{12}{}^3 s_{14} 
  }{2 s_{13}{}^4}
  A^{(0)}
  \left(
  2 s_{12} s_{14} \omNfpNs
  -\fmNf s_{13}{}^2
  \right)
\end{align}

\subsubsection{The $-++-$ Helicity Amplitude}

The integrands for this configuration are:
\begin{align}
  &\Delta_{7;12*34*}^{\dbox, {\rm ns},-++-} = -s_{14} s_{12}{}^2 \nonumber\\&
  -\frac{4 s_{12}{}^2 \omNfpNs 
  }{s_{14}{}^3}
  \Big(
  \MP{k}{4}^3 \left(8 \MP{q}{1}-4 s_{14}\right)
  +\MP{k}{4}^2 \left(s_{14}{}^2-2 \left(3 s_{12}+4
  s_{14}\right) \MP{q}{1}\right)
  \nonumber\\&
  +\MP{k}{4} \MP{q}{1} \left(4 \MP{q}{1}-3 s_{12}-2 s_{14}\right) \left(2 \MP{q}{1}-s_{14}\right)
  +4 \MP{k}{4}^4+\MP{q}{1}^2 \left(s_{14}-2 \MP{q}{1}\right){}^2
  \Big)
  \nonumber\\&
  +\frac{2 \fmNf \tmNs s_{12}{}^2
  }{s_{14}{}^2} 
  \MP{k}{4}
  \MP{q}{1} \left(2 (\MP{k}{4}+\MP{q}{1})-s_{14}\right)
  \nonumber\\&
  +\frac{\fmNf s_{12}{}^2 
  }{s_{14}{}^2}
  \Big(
  s_{14}{}^2
  (\MP{k}{4}+\MP{q}{1})-2 s_{14} \left(-\MP{k}{4} \MP{q}{1}+\MP{k}{4}^2+\MP{q}{1}^2\right)
  \nonumber\\&
  -8 \MP{k}{4} \MP{q}{1} (\MP{k}{4}+\MP{q}{1})
  \Big)
\end{align}
\begin{align}
  &\Delta_{7;12*34*}^{\dbox, {\rm s},-++-} = \nonumber\\&
  \frac{\fmNf \tmNs s_{12}{}^2 
  }{2 s_{13} s_{14}{}^2}
  \Big(
  s_{12} s_{14}{}^2 (\MP{k}{\w}+\MP{q}{\w})
  -2 s_{12} s_{14} (\MP{k}{4} \MP{q}{\w}+\MP{q}{1} \MP{k}{\w})
  \nonumber\\&
  +8 s_{13} \MP{k}{4} \MP{q}{1} (\MP{k}{\w}+\MP{q}{\w})
  \Big)
  \nonumber\\&
  +\frac{s_{12}{}^2 \omNfpNs
  }{s_{13} s_{14}{}^3}
  \Big(
  2 s_{12} s_{14}{}^3 (\MP{k}{\w}+\MP{q}{\w})
  \nonumber\\&
  +s_{12} s_{14}{}^2 \left(3 s_{12} (\MP{k}{\w}+\MP{q}{\w})-4 \MP{k}{4} \MP{q}{\w}-4
  \MP{q}{1} \MP{k}{\w}\right) 
  \nonumber\\&
  +2 s_{14} \big(
  4 s_{13} \left(
  2 \MP{k}{4} \MP{q}{1} (\MP{k}{\w}+\MP{q}{\w})+\MP{k}{4}^2
  \MP{k}{\w}+\MP{q}{1}^2 \MP{q}{\w}\right)
  \nonumber\\&
  -3 s_{12}{}^2 (\MP{k}{4} \MP{q}{\w}+\MP{q}{1} \MP{k}{\w})
  \big)
  -8 s_{13} \big(
  -3 s_{12}\MP{k}{4} \MP{q}{1} (\MP{k}{\w}+\MP{q}{\w})
  \nonumber\\&
  +2 \left(\MP{k}{4}^2 (\MP{k}{4}+2 \MP{q}{1}) \MP{k}{\w}+\MP{q}{1}^2 (2\MP{k}{4}+\MP{q}{1}) \MP{q}{\w}\right)
  \big)
  \Big)
  \nonumber\\&
  +\frac{\fmNf s_{12}{}^2 
  }{2 s_{13} s_{14}{}^2}
  \Big(
  -3 s_{12} s_{14}{}^2 (\MP{k}{\w}+\MP{q}{\w})
  -16 s_{13} \MP{k}{4} \MP{q}{1} (\MP{k}{\w}+\MP{q}{\w})
  \nonumber\\&
  +2 s_{14} \left(3 s_{12} (\MP{k}{4} \MP{q}{\w}+\MP{q}{1} \MP{k}{\w})-2 s_{13} (\MP{k}{4} \MP{k}{\w}+\MP{q}{1} \MP{q}{\w})\right)
  \Big)
\end{align}
which lead to coefficients of the master integrals of,
\begin{align}
  C_1^{-++-} &=
  -\frac{s_{12}{}^2 
  }{4 s_{14}{}^2}
  A^{(0)} 
  \big(
  2 s_{12} \left(10 s_{12}{}^2+11 s_{14} s_{12}+2 s_{14}{}^2\right) \omNfpNs \nonumber\\&
  +s_{14} \left(\fmNf \tmNs s_{12} \left(2 s_{12}+s_{14}\right)-\fmNf s_{12} \left(4 s_{12}+s_{14}\right)+4 s_{14}{}^2\right)
  \big)
  \\
  C_2^{-++-} &= 
  \frac{3 s_{12}{}^3 
  }{2 s_{14}{}^3}
  A^{(0)}
  \big(
  \left(20 s_{12}{}^2+22 s_{14} s_{12}+4 s_{14}{}^2\right) \omNfpNs \nonumber\\&
  +s_{14} \left(\fmNf \tmNs \left(2 s_{12}+s_{14}\right)-\fmNf \left(4
   s_{12}+s_{14}\right)\right)
  \big)
\end{align}
We notice that the explicit expressions for all helicity amplitudes never contain tensor coefficients higher than
rank four. In other words the coefficients $c_{410}$, $c_{140}$, $c_{311}$, and $c_{132}$ are zero even in
pure Yang-Mills though we were not able to exclude them \textit{a priori} from the renormalization
constraints.

\subsection{Non-Planar Crossed Box}

\subsubsection{The $--++$ Helicity Amplitude}

The integrands for this configuration are:
\begin{align}
  &\Delta_{7;1*34*2}^{\xbox, {\rm ns},--++} =
  -s_{14} s_{12}{}^2 \nonumber\\&
  +\frac{1}{2} \fmNf s_{14} \left(-2 \left(s_{12}+2 s_{14}\right) \MP{q}{2}-4\MP{q}{2}^2+s_{13} s_{14}\right)
  \nonumber\\&
  +\frac{ s_{14} \omNfpNs }{ s_{12}{}^2 } 
  \Big(
  -4 \MP{q}{2}^2 \left(2 \MP{q}{2}+s_{12}\right){}^2
  -s_{14}{}^2 \left(-8 s_{13} \MP{q}{2}+24 \MP{q}{2}^2+s_{13}{}^2\right) \nonumber\\&
  +4 s_{14} \MP{q}{2} \left(-6 s_{12} \MP{q}{2}-8 \MP{q}{2}^2+s_{12} s_{13}\right)
  \Big)
\end{align}
\begin{align}
  &\Delta_{7;1*34*2}^{\xbox, {\rm s},--++} = \nonumber\\&
  -\frac{ \fmNf s_{12} }{ 2 s_{13} }
  \left(\MP{q}{\omega} \left(s_{13} \left(2 \MP{q}{2}+s_{14}\right)-2 s_{12} \MP{k}{3}\right)+2
  \MP{k}{\omega} \left(\left(s_{12}+2 s_{14}\right) \MP{q}{2}-s_{13} s_{14}\right)\right)
  \nonumber\\&
  +\frac{ s_{14} \omNfpNs }{ s_{12}{}^2 }
  \Big(
  2 \MP{q}{2} \left(s_{12}{}^2 (3 \MP{k}{\omega}-\MP{q}{\omega})+3 s_{14} s_{12} (2 \MP{k}{\omega}+\MP{q}{\omega})+6
  s_{14}{}^2 \MP{q}{\omega}\right) \nonumber\\&
  +16 \left(s_{12}+2 s_{14}\right) \MP{q}{2}^2 \MP{q}{\omega}+16 \MP{q}{2}^3 \MP{q}{\omega}
  +s_{12} s_{13} s_{14}(\MP{q}{\omega}-2 \MP{k}{\omega}) \nonumber\\&
  +2 \MP{k}{3} \left(16 \left(s_{12}+2 s_{14}\right) \MP{q}{2}+16 \MP{q}{2}^2+s_{12}{}^2+12 s_{14}{}^2+12 s_{12} s_{14}\right) \MP{q}{\omega}
  \Big)
\end{align}
After applying the IBP relations we obtain the following coefficients of the master integrals via eq.
\eqref{eq:xboxMIcoeffs},
\begin{align}
  C_1^{--++} &= 
  \frac{1}{4} s_{14} A^{(0)}\left(-\frac{2 s_{13}{}^2 s_{14}{}^2 \omNfpNs}{s_{12}{}^2}+\fmNf s_{13} s_{14}-4 s_{12}{}^2\right)
  \\
  C_2^{--++} &=
  -\frac{s_{14}}{2 s_{12}{}^2} A^{(0)} \left(s_{13}-s_{14}\right) \left((4-n_f) s_{12}{}^2-2 s_{13} s_{14} (1-n_f+n_s)\right)
\end{align}

\subsubsection{The $-+-+$ Helicity Amplitude}

The integrands for this configuration are:
\begin{align}
  &\Delta_{7;1*34*2}^{\xbox, {\rm ns},-+-+} =
  -s_{14} s_{12}{}^2 \nonumber\\& 
  +\frac{\fmNf s_{14} s_{12}{}^2 }{s_{13}{}^3}
  \Big(
  -2 s_{13} \left(3 \MP{k}{3} \MP{q}{2}+3 \MP{k}{3}^2+\MP{q}{2}^2\right) \nonumber\\&
  +8 \MP{k}{3}\MP{q}{2}(\MP{k}{3}+\MP{q}{2})+s_{13}{}^2 \MP{q}{2}
  \Big) \nonumber\\&
  - \frac{2 \fmNf \tmNs s_{14} s_{12}{}^2 }{s_{13}{}^3}
  \MP{k}{3} (\MP{k}{3}+\MP{q}{2}) \left(2 \MP{q}{2}-s_{13}\right) \nonumber\\&
  + \frac{4 s_{14} s_{12}{}^2 \omNfpNs }{s_{13}{}^4}
  \Big(
  2 s_{14} \MP{k}{3} (\MP{k}{3}+\MP{q}{2}) \left(2 \MP{q}{2}-s_{13}\right)
  -8 \MP{k}{3}^3 \MP{q}{2} \nonumber\\&
  -4 \MP{k}{3}^2 \MP{q}{2}^2-4 \MP{k}{3}^4
  -s_{13}{}^2 \MP{q}{2}^2+4 s_{13}\MP{q}{2}^3-4 \MP{q}{2}^4
  \Big)
\end{align}
\begin{align}
  &\Delta_{7;1*34*2}^{\xbox, {\rm s},-+-+} = \nonumber\\&
  \frac{\fmNf s_{12} s_{14}}{2 s_{13}{}^3}
  \Big(
  2 s_{12} \MP{k}{\omega} \left(3 s_{13} (2 \MP{k}{3}+\MP{q}{2})-8 \MP{k}{3} \MP{q}{2}\right)
  \nonumber\\&
  +\MP{q}{\omega} \left(2 s_{12} s_{13} \left(-3\MP{k}{3}-5 \MP{q}{2}+s_{13}\right)+s_{14} \left(8 \MP{k}{3}+3 s_{13}\right)\left(s_{13}-2 \MP{q}{2}\right)\right)
  \Big) \nonumber\\&
  -\frac{\fmNf \tmNs s_{12} s_{14}}{2s_{13}{}^3}
  \Big(
  2 s_{12} \MP{k}{\omega} \left(s_{13} (2 \MP{k}{3}+\MP{q}{2})-4 \MP{k}{3}\MP{q}{2}\right)
  \nonumber\\&
  +\MP{q}{\omega} \left(2 \MP{k}{3} \left(2 s_{14} \left(s_{13}-2 \MP{q}{2}\right)-s_{12} s_{13}\right)+s_{13}{}^2 \left(2 \MP{q}{2}+s_{14}\right)\right)
  \Big) \nonumber\\&
  +\frac{s_{14} \omNfpNs}{s_{13}{}^4}
  \Big(
  2 s_{12} \MP{k}{\omega} 
  \big(
  8 s_{12} \MP{k}{3}^2 (\MP{k}{3}+\MP{q}{2}) \nonumber\\&
  +s_{14}{}^2 \left(4 \MP{k}{3} \MP{q}{2}-s_{13}(2\MP{k}{3}+\MP{q}{2})\right) \nonumber\\&
  +2 s_{14} \left(s_{13} \left(s_{12} (2 \MP{k}{3}+\MP{q}{2})+2 \MP{k}{3}^2\right)-4 \MP{k}{3} \MP{q}{2}\left(\MP{k}{3}+s_{12}\right)\right)
  \big) \nonumber\\&
  +\MP{q}{\omega}
  \big(
  s_{14}{}^3 \left(-\left(8 \MP{k}{3} \left(s_{13}-3 \MP{q}{2}\right)+s_{13}{}^2\right)\right)
  \nonumber\\&
  +2 s_{14}{}^2 \left(\MP{k}{3} \left(8\MP{q}{2}^2+3 s_{12} s_{13}\right)+s_{13}{}^2
  \left(s_{12}-\MP{q}{2}\right)\right)
  \nonumber\\&
  +4 s_{12} s_{13} s_{14} \left(s_{13} \MP{q}{2}-s_{12} \MP{k}{3}\right)-4 s_{12}{}^2 \MP{q}{2} \left(s_{13}-2\MP{q}{2}\right){}^2
  \big)
  \Big) %
\end{align}
which lead to coefficients of the master integrals of,
\begin{align}
  C_1^{-+-+} &=
  \frac{s_{12}{}^2 s_{14}}{4 s_{13}{}^3}  A^{(0)}\left(2 s_{12} s_{14}{}^2 \omNfpNs-\fmNf s_{14} s_{13}{}^2-4 s_{13}{}^3\right)
  \\
  C_2^{-+-+} &=
  \frac{s_{12}{}^2 s_{14}}{2 s_{13}{}^4}  A^{(0)}\left(s_{13}+3 s_{14}\right) \left(2 s_{12} s_{14} \omNfpNs-\fmNf s_{13}{}^2\right)
\end{align}

\subsubsection{The $-++-$ Helicity Amplitude}

The integrands for this configuration are:
\begin{align}
  &\Delta_{7;1*34*2}^{\xbox, {\rm ns},-++-} = -s_{14} s_{12}{}^2\nonumber\\&
  \frac{2 (4-n_f) (3-n_s) s_{12}{}^2 }{s_{14}{}^2}
  \MP{k}{3} (\MP{k}{3}+\MP{q}{2}) \left(2 \MP{q}{2}+s_{14}\right)
  \nonumber\\&
  -\frac{4 s_{12}{}^2 (1-n_f+n_s) 
  }{s_{14}{}^3}
  \Big(
  2 \MP{k}{3}^2 \left( \MP{q}{2} \left(\MP{q}{2}+s_{13}\right)+s_{13} s_{14}\right)
  \nonumber\\&
  +2 s_{13} \MP{k}{3} \MP{q}{2} \left(2 \MP{q}{2}+s_{14}\right)
  +8 \MP{k}{3}^3 \MP{q}{2}+4 \MP{k}{3}^4+\MP{q}{2}^2
  \left(2 \MP{q}{2}+s_{14}\right){}^2
  \Big)
  \nonumber\\&
  -\frac{(4-n_f) s_{12}{}^2 }{s_{14}{}^2}
  \Big(
  2 s_{14} \left(3 \MP{k}{3} \MP{q}{2}+3 \MP{k}{3}^2+\MP{q}{2}^2\right)
  \nonumber\\&
  +8 \MP{k}{3} \MP{q}{2} (\MP{k}{3}+\MP{q}{2})+s_{14}{}^2 \MP{q}{2}
  \Big)
\end{align}
\begin{align}
  &\Delta_{7;1*34*2}^{\xbox, {\rm s},-++-} = \nonumber\\&
  \frac{(1-n_f+n_s) 
  }{s_{14}{}^3}
  \Big(
  2 s_{12} \MP{k}{\w} \big(
  s_{14}{}^2 \left(4 \MP{k}{3} \MP{q}{2}-s_{13} (2 \MP{k}{3}+\MP{q}{2})\right)
  \nonumber\\&
  +2 s_{14} \left(s_{13} \left(s_{12} (2 \MP{k}{3}+\MP{q}{2})+2 \MP{k}{3}^2\right)-4 \MP{k}{3} \MP{q}{2} \left(\MP{k}{3}+s_{12}\right)\right)
  \nonumber\\&
  +8 s_{12} \MP{k}{3} \left(\MP{k}{3} (\MP{k}{3}+\MP{q}{2})-s_{12} \MP{q}{2}\right)
  \big)
  \nonumber\\&
  +\MP{q}{\w} \big(
  s_{14}{}^3 \left(8 \MP{k}{3} \left(3 \MP{q}{2}-s_{13}\right)-s_{13}{}^2\right)
  \nonumber\\&
  +2 s_{14}{}^2 \left(\MP{k}{3} \left(8 \MP{q}{2}^2+3 s_{12} s_{13}\right)+s_{12} \left(5 s_{12} \MP{q}{2}+s_{13}{}^2\right)\right)
  \nonumber\\&
  +4 s_{12}{}^2 s_{14} \left(\MP{q}{2} \left(-4 \MP{k}{3}+4 \MP{q}{2}+s_{12}\right)+s_{13} \MP{k}{3}\right)
  -2 s_{14}{}^4 \MP{q}{2}+16 s_{12}{}^2 \MP{q}{2}^3
  \big)
  \Big)
  \nonumber\\&
  -\frac{(4-n_f) (3-n_s) s_{12} 
  }{2 s_{14}{}^2}
  \Big(
  2 s_{12} \MP{k}{\w} \left( s_{14} (2 \MP{k}{3}+\MP{q}{2})+4 \MP{k}{3}\MP{q}{2}\right)
  \nonumber\\&
  +s_{14} \MP{q}{\w} \left(2 \MP{k}{3} \left(4 \MP{q}{2}+s_{12}+2 s_{14}\right)+s_{13} \left(2 \MP{q}{2}+s_{14}\right)\right)
  \Big)
  \nonumber\\&
  +\frac{(4-n_f) s_{12} 
  }{2 s_{14}{}^2}
  \Big(
  2 s_{12} \MP{k}{\w} \left(3 s_{14} (2 \MP{k}{3}+\MP{q}{2})+8 \MP{k}{3} \MP{q}{2}\right)
  \nonumber\\&
  -s_{14} \MP{q}{\w} \left(\left(s_{12}+3 s_{14}\right) \left(2 \MP{q}{2}+s_{14}\right)-2 \MP{k}{3}\left(8 \MP{q}{2}+s_{12}+4 s_{14}\right)\right)
  \Big)
\end{align}
which lead to coefficients of the master integrals of,
\begin{align}
  C_1^{-++-} &=
  \frac{s_{12}{}^2 }{4 s_{14}{}^2} A^{(0)}
  \left(2 s_{12} s_{13}{}^2 (1-n_f+n_s)-s_{14}{}^2 \left((4-n_f) s_{13}+4 s_{14}\right)\right)
  \\
  C_2^{-++-} &=
  \frac{s_{12}{}^2 }{2 s_{14}{}^3} A^{(0)}
  \left(3 s_{13}+s_{14}\right) \left((4-n_f) s_{14}{}^2-2 s_{12} s_{13} (1-n_f+n_s)\right)
\end{align}

As in the planar double box, non-zero tensor coefficients never appear higher than
rank four since the values of the coefficient $c_{050}$, $c_{320}$, $c_{410}$, $c_{060}$, $c_{420}$, $c_{052}$, or $c_{142}$ 
turn out be zero independent of the helicity configuration.

\subsection{Penta-Box}

This topology only appears in helicity configurations which are zero at tree-level. There are two
independent contributions, both of which vanish in super-symmetric theories.

\subsubsection{The $-+--$ Helicity Amplitude}

\begin{align}
  &\Delta_{7;1*234*}^{\pbox,ns,-+--} =
  4i \omNfpNs
  \nonumber\\&\times 
  \frac{\A{13}^2\A{14}^2s_{12}^3}{\A{12}^2s_{13}^2s_{14}}
  \left( \MP{k}{2}\MP{k}{4}-\frac{2}{s_{12}}\MP{k}{2}^2\MP{k}{4}+\frac{2}{s_{14}}\MP{k}{2}\MP{k}{4}^2 \right)
  \\
  &\Delta_{7;1*234*}^{\pbox,s,-+--} =
  -8i \omNfpNs
  \nonumber\\&\times 
  \frac{\A{13}^2\A{14}^2s_{12}^3}{\A{12}^2s_{13}^2s_{14}^2}
  \left( \MP{k}{2}\MP{k}{4}-\frac{2}{s_{12}}\MP{k}{2}^2\MP{k}{4}+\frac{2}{s_{14}}\MP{k}{2}\MP{k}{4}^2 \right)
  \MP{q}{\w}
\end{align}

\subsubsection{The $--+-$ Helicity Amplitude}

\begin{align}
  &\Delta_{7;1*234*}^{\pbox,ns,--+-} = 
  4i \omNfpNs
  \nonumber\\&\times 
  \frac{\A{12}^2\A{14}^2s_{13}^2}{\A{13}^2s_{12}s_{14}}
  \left( \MP{k}{2}\MP{k}{4}-\frac{2}{s_{12}}\MP{k}{2}^2\MP{k}{4}+\frac{2}{s_{14}}\MP{k}{2}\MP{k}{4}^2 \right)
  \\
  &\Delta_{7;1*234*}^{\pbox,s,--+-} =
  -8i \omNfpNs
  \nonumber\\&\times 
  \frac{\A{12}^2\A{14}^2s_{13}^2}{\A{13}^2s_{12}s_{14}^2}
  \left( \MP{k}{2}\MP{k}{4}-\frac{2}{s_{12}}\MP{k}{2}^2\MP{k}{4}+\frac{2}{s_{14}}\MP{k}{2}\MP{k}{4}^2 \right)
  \MP{q}{\w}
\end{align}

\section{Conclusions}

In recent years, unitarity methods have been particularly useful in the computation of multi-loop
scattering amplitudes in super-symmetric gauge theories and gravity. In this paper we considered
the possibility of computing two-loop scattering amplitudes in a general renormalizable gauge theory
with no super-symmetry via generalised unitarity cuts.

The traditional unitarity approach to one-loop amplitudes relies on knowing a basis of scalar
integrals in advance of the computation. Since such a basis is not known at two-loops, we looked
to Gram matrix identities to constrain the general form of the integrand. This polynomial
form can then be efficiently fitted by systematically evaluating products of tree-level amplitudes
over a complete set of complex on-shell solutions to the loop momentum cut constraints. We derived a
general map between the expansion of the tree level input and the coefficients of the integrand
using only elementary linear algebra.

The general integrand can be reduced to a set of master integrals using well known integration by
parts identities. Using such identities we have derived master formulae for the three independent
seven propagator topologies for $2\to2$ scattering. The method applies equally well to planar
and non-planar topologies.

As a test of our approach we computed the hepta-cut part of two-loop helicity amplitudes in
Yang-Mills theory with adjoint fermion and scalars. This allowed us to check our results against the
known results in super-symmetric Yang-Mills theories.

Though a small step towards the complete reduction of an arbitrary two-loop amplitude, we hope the
Gram matrix method introduced here will be of use in studying both $D$-dimensional cuts and cuts with
fewer propagators. The extension to treat amplitudes with a higher number of external legs should
also be possible following the basic steps described here, nevertheless a large number of basic
topologies would be required.

Another interesting direction would be the application of the technique to higher loop amplitudes.
Though the solution to the Gram constraints will certainly be much more involved, the basic procedure
for parameterising the integrand would be expected to apply.

\acknowledgments{%
We are grateful to Pierpaolo Mastrolia, Giovanni Ossola, Nigel Glover and Kasper Larsen for useful discussions. We
are especially indebted to Zvi Bern for providing analytic expressions for the results of
\cite{Bern:2002tk} in terms of master integrals used to check the results in sec. \ref{sec:4gamps}.
YZ acknowledges support from the Danish Council for Independent Research – Natural Sciences (FNU)
grant number 11-107241.
}

\appendix

\section{Conventions for Spinors and Spinor Products \label{app:spinors}}

Throughout this paper we will make use of the well known spinor-helicity formalism in
four-dimensions. A massless vector, $p$, can be written as,
\begin{equation}
  p^\mu = \frac{1}{2} \spAB{p}{\gamma^\mu}{p}.
\end{equation}
where are $\la p|$ and $|p]$ two component Weyl spinors of negative and positive helicity respectively.
Helicity amplitudes can then be written in terms of spinor products $\A{pq}$ and $\B{pq}$ where
$s_{pq} = (p+q)^2 = \A{pq}\B{qp}$. We have also made use of the massless
decomposition of two massive vectors, $P_1$ and $P_2$,  in a basis of two massless vectors, $\kf1,\kf2$,
\begin{align}
  \kfm1 &= \frac{\gamma_{12} \left(\gamma_{12} P_1^\mu - P_1^2P_2^\mu
  \right)}{\gamma_{12}^2-P_1^2P_2^2}\\
  \kfm2 &= \frac{\gamma_{12} \left(\gamma_{12} P_2^\mu - P_2^2P_1^\mu
  \right)}{\gamma_{12}^2-P_1^2P_2^2}\\
  \gamma_{12} &= \vec{P}_1\cdot \vec{P}_2 + \text{sign}(\vec{P}_1\cdot \vec{P}_2)\sqrt{(\vec{P}_1\cdot \vec{P}_2)^2-P_1^2P_2^2}
  \label{eq:kflatbasis}
\end{align}
where we choose the sign in front of the square root to ensure that $\gamma_{12}>0$. Using these
definition the size of the spurious vector in eq. \eqref{eq:D4spuriousvector} is,
\begin{equation}
 \w^2=-\frac{\det G(P_1,P_2,P_3)}{\det G(P_1,P_2)}.
\end{equation}

\section{Tree Level Amplitudes \label{app:trees}}

For completeness we list the well known formula for the independent tree level helicity amplitudes
used in this paper:
\begin{align}
  A^{(0)}(1^-,2^-,3^+) &= i\frac{\A{12}^3}{\A{23}\A{31}} \\
  A^{(0)}(1_q^-,2^-,3_{\qb}^+) &= i\frac{\A{12}^2}{\A{31}} \\
  A^{(0)}(1_q^+,2^-,3_{\qb}^-) &= -i\frac{\A{23}^2}{\A{31}} \\ 
  A^{(0)}(1_s,2^-,3_s) &= i\frac{\A{12}\A{23}}{\A{31}} \\
  A^{(0)}(1_s,2^-,3_s) &= i\frac{\A{12}\A{23}}{\A{31}} \\
  A^{(0)}(1_q^-,2_q^-,3_s) &= i\A{12} \\
  A^{(0)}(1_{\qb}^-,2_{\qb}^-,3_s) &= i\A{12}
  \label{eq:3amps}
\end{align}
the \MHVb amplitudes are obtained by complex conjugation $(\A{\,} \leftrightarrow -\B{\,})$.
All other amplitudes not related by parity or cyclic symmetries are zero.

The non-zero four-gluon amplitudes used in section \ref{sec:4gamps} are:
\begin{align}
  A^{(0)}(1^-,2^-,3^+,4^+) &= i\frac{\A{12}^4}{\A{12}\A{23}\A{34}\A{41}} \\
  A^{(0)}(1^-,2^+,3^+,4^-) &= i\frac{\A{14}^4}{\A{12}\A{23}\A{34}\A{41}} \\
  A^{(0)}(1^-,2^+,3^-,4^+) &= i\frac{\A{13}^4}{\A{12}\A{23}\A{34}\A{41}}
  \label{eq:4gamps}
\end{align}

\section{General Solution for the One-Loop Box \label{app:1lbox}}

Just the for the purposes of completeness we give the explicit expressions for the
terms used in section \ref{sec:1lD4}. We take the basis $\{P_1,P_2,P_3,\w\}$ to span our
space as before. For arbitrary kinematics with massless propagators the solutions are:
\begin{align}
  k^{(1),\mu} = \frac{1}{\gamma^2-P_1^2P_2^2} \Big(&
   P_2^2\left( \gamma_{12}+P_1^2 \right) \kfm1
  -P_1^2\left( \gamma_{12}+P_2^2 \right) \kfm2 \nonumber\\&
  -\frac{C+\sqrt{D}}{4\spAB{\kf1}{P_3}{\kf2}} \spAB{\kf1}{\gamma^\mu}{\kf2}
  -\frac{C-\sqrt{D}}{4\spAB{\kf2}{P_3}{\kf1}} \spAB{\kf2}{\gamma^\mu}{\kf1}
  \Big) \\
  k^{(2),\mu} = \frac{1}{\gamma^2-P_1^2P_2^2} \Big(&
   P_2^2\left( \gamma_{12}+P_1^2 \right) \kfm1
  -P_1^2\left( \gamma_{12}+P_2^2 \right) \kfm2 \nonumber\\&
  -\frac{C-\sqrt{D}}{4\spAB{\kf1}{P_3}{\kf2}} \spAB{\kf1}{\gamma^\mu}{\kf2}
  -\frac{C+\sqrt{D}}{4\spAB{\kf2}{P_3}{\kf1}} \spAB{\kf2}{\gamma^\mu}{\kf1}
  \Big)
\end{align}
where the massless vectors $\kf1,\kf1$ are defined in Appendix \ref{app:spinors}.
The constant $C$ is
\begin{align}
  C =& 
   2P_2^2\left( \gamma_{12}+P_1^2 \right)(\kf1\cdot P_3)
  -2P_1^2\left( \gamma_{12}+P_2^2 \right)(\kf2\cdot P_3) \nonumber\\&
  -\left( \gamma_{12}^2-P_1^2P_2^2 \right)\left( P_3^2+2(P_2\cdot P_3) \right)
  \label{eq:C}
\end{align}
while $D$ can be expressed as:
\begin{equation}
  D = C^2-4P_1^2P_2^2\spAB{\kf1}{P_3}{\kf2}\spAB{\kf2}{P_3}{\kf1}\left( \gamma_{12}+P_1^2 \right)\left( \gamma_{12}+P_2^2 \right)
  \label{eq:D}
\end{equation}
which is related to the normalisation of the spurious coefficient $c_{2}$ by,
\begin{equation}
  \sqrt{V_4} = -\left( \gamma_{12}^2-P_1^2P_2^2 \right)\sqrt{D}.
  \label{eq:rtV4}
\end{equation}

\section{Notation for the Two-Loop Integrands \label{app:notation}}

\begin{figure}[h]
  \begin{center}
    \psfrag{i1}{\small$i_1$}
    \psfrag{ijm1}{\small$i_{r-1}$}
    \psfrag{ij}{\small$i_r$}
    \psfrag{ij1}{\small$i_{r+1}$}
    \psfrag{ikm1}{\small$i_{s-1}$}
    \psfrag{ik}{\small$i_s$}
    \psfrag{ik1}{\small$j_1$}
    \psfrag{ik2}{\small$j_2$}
    \psfrag{inm1}{\small$j_{t-1}$}
    \psfrag{in}{\small$j_t$}
    \psfrag{1}{$1$}
    \psfrag{2}{$2$}
    \psfrag{3}{$3$}
    \psfrag{4}{$4$}
    \psfrag{5}{$5$}
    \psfrag{6}{$6$}
    \psfrag{7}{$7$}
    \psfrag{a}{$(a)$}
    \psfrag{b}{$(b)$}
    \psfrag{c}{$(c)$}
    \includegraphics[width=\textwidth]{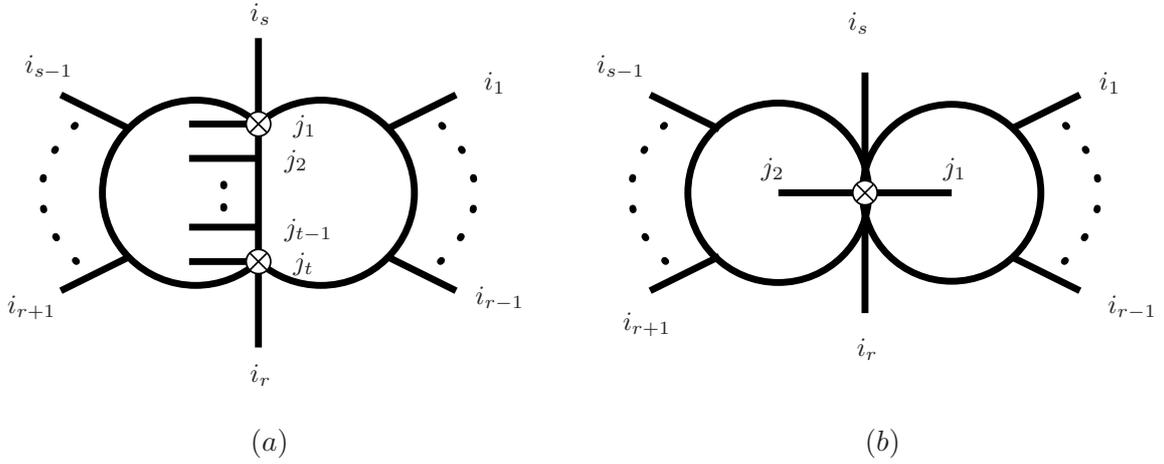}
  \end{center}
  \caption{
  Notation for a general ordered two loop amplitude. (a) A topology with two intersections and (b) 
  A topology with a single intersection.} 
  \label{fig:parent}
\end{figure}

An $n$-point two-loop primitive amplitude can be described by an ordered 
set of momenta on external lines, say $S=\{p_1,\cdots,p_m\}$ and another set on
internal lines, $T=\{p_{m+1},\cdots,p_n\}$.

We label the integrands by a set of indices, $\{i_x\}$, corresponding to the position in the set $S$ and a set
of indices, $\{j_x\}$, for the position the set $T$. The momentum leaving each vertex
$x=1,\ldots,t$ is given by:
\begin{equation}
  P_x = \sum_{a=i_x}^{i_{x+1}-1} S_a,
\end{equation}
where the sum is considered to be cyclic modulo $m$. The momentum leaving each vertex $y=t+1,\ldots,s$
are given by:
\begin{equation}
  P_y = \sum_{a=j_k}^{j_{k+1}-1} T_a,
\end{equation}
where $j_{t+1}-1=n$.

The vertices with intersections between the loops require a slightly more elaborate notation. Each
intersection is labeled by a set of indices give the ranges of momenta entering at each
section, $[i_x,j_y]$.

The integrand functions which we call $\Delta(k,q)$, are given a subscript according to the indices
above with the number of cut propagators as a prefix. The double loop topology shown in figure
\ref{fig:parent}(a) would be represented as,
\begin{equation}
  \Delta_{s+t-1;i_1\cdots i_{r-1}[i_{r},j_t]i_{r+1}\cdots i_{s-1}[i_s,j_1]j_2\cdots j_{t-1}}.
  \label{eq:not.2vert}
\end{equation}
In the case where the intersection vertex has no external legs attached, we represent it with a
`$*$':
\begin{equation}
  \Delta_{s+t+3;i_1\cdots i_r*i_{r+1}\cdots i_s*j_1\cdots j_t}.
  \label{eq:not.2vert*}
\end{equation}
Following the same structure the butterfly topology, shown in fig. \ref{fig:parent}(b), would be
represented as,
\begin{equation}
  \Delta_{s;i_1\cdots i_{r-1}[i_r,i_s,j_1,j_2]i_{r+1}\cdots i_{s-1}}
  \label{eq:not.1vert}
\end{equation}
We note that though the set of indices uniquely defines a topology, it does not account for possible
symmetries between primitive amplitudes. The butterfly and higher multiplicity double loop topologies
are of course beyond the scope of this paper.

\providecommand{\href}[2]{#2}\begingroup\raggedright

\endgroup

\end{document}